\documentclass{ar-1col-S2O}

\usepackage[numbers,compress,square]{natbib}
\usepackage{url}
\usepackage{subfigure}
\usepackage{graphicx}
\usepackage{amsmath,amssymb}
\usepackage{aas_macro}

\makeatletter
\renewcommand{\@thesubfigure}{\hskip\subfiglabelskip}
\makeatother

\jname{Annu. Rev. Nucl. Part. Sci}
\jvol{74}
\jyear{2024}
\doi{10.1146/((annurev-nucl-102020-124444))}

\begin{document}

% Page header
\markboth{Zhang}{Multi-wavelength and Multi-messenger Counterparts of FRBs}

% Title
\title{Multi-wavelength and Multi-messenger Counterparts of Fast Radio Bursts}

%Authors, affiliations address.
\author{Bing Zhang$^{1,2}$
\affil{$^1$The Nevada Center for Astrophysics, University of Nevada, Las Vegas; email: bing.zhang@unlv.edu}
\affil{$^2$Department of Physics and Astronomy, University of Nevada, Las Vegas}
}

%Abstract
\begin{abstract}
Fast radio bursts are brief, highly dispersed bursts detected in the radio band, originating from cosmological distances. The only such event detected in the Milky Way galaxy, FRB 20200428DD, was associated with an X-ray burst emitted by a magnetar named SGR J1935+2154, revealing the first case of a multi-wavelength counterpart of an FRB. Counterparts in other wavelengths accompanying or following FRBs, as well as the bright emission associated with the progenitor of the FRB engine, have been proposed in various FRB models, but no robust detection has been made so far. In general, FRBs as we know them are not favorite multi-messenger emitters. Nonetheless, possible neutrino and gravitational wave emission signals associated with FRBs or FRB-like events have been discussed in the literature. Here I review these suggested multi-wavelength and multi-messenger counterparts of FRBs or FRB-like events and the observational progress in searching for these signals. The topics include multi-wavelength (X-rays, $\gamma$-rays, optical) emission and neutrino emission from FRBs within the framework of the magnetar source models and possible FRB-like events associated with gravitational waves.
\end{abstract}

%Keywords, etc.

\maketitle

%Table of Contents
\tableofcontents

% Heading 1

\section{Introduction}

Fast radio bursts (FRBs) \cite{lorimer07,thornton13} are millisecond-duration radio bursts originating from cosmological distances. Compared with radio pulsars which held the record of brightness temperature in the universe before their discovery, FRBs boost the record by $\sim$ 10 orders of magnitude and reached up to $\sim 10^{36}$ K, which signifies the most luminous radio emission process in the universe by a hitherto unidentified coherent emission mechanism. On the other hand, compared with the huge luminosities (typically $\sim 10^{52} \ {\rm erg \ s^{-1}}$) of gamma-ray bursts (GRBs), the FRB luminosities (typically $\sim 10^{41} \ {\rm erg \ s^{-1}}$) are minuscule. As a result, FRBs are not particularly attractive high-energy or multi-messenger emitters. Nonetheless, a megaJansky burst named FRB 20200428D detected by the Canadian Hydrogen Intensity Mapping Experiment (CHIME, two pulses detected) \cite{CHIME-SGR} and the Transient Astronomical Radio Emission 2 (STARE2, only one pulse detected) \cite{STARE2-SGR} was associated with a hard X-ray / soft $\gamma$-ray burst detected by multiple high-energy observatories: Insight-HXMT \cite{HXMT-SGR}, Integral \cite{Integral-SGR}, Konus/Wind \cite{Konus-SGR}, and AGILE \cite{AGILE-SGR}. The source of the emissions is a Galactic magnetar named SGR J1935+2154. This watershed event presents the first (and the only one so far) multi-wavelength counterpart of an FRB, thanks to its close distance to Earth as the only FRB detected within the Milky Way galaxy. For cosmological events, despite intense searches of counterparts in other wavelengths, no positive detection has been made. 

Thus, writing a review article on multi-wavelength (especially high-energy) and multi-messenger emission from FRBs, a subject of the interest of the ARNPS readers, is not an easy task. One can only present some observational upper limits and theoretical {predictions} (if the physical setup is confidently known) or {speculations} (if the physical setup itself is uncertain) on various multi-wavelength and multi-messenger emission signals from FRBs.  Nonetheless, because the nature of cosmological FRBs is largely unknown, these predictions/speculations and upper limits are worth summarizing at this stage. This is the purpose of this review. A brief summary of the FRB field is presented in \S\ref{sec:FRBs}. The proposed multi-wavelength electromagnetic counterparts and their observational constraints are presented in \S\ref{sec:MW}. This is followed by the discussion of the neutrino signals in \S\ref{sec:nus} and gravitational wave signals in \S\ref{sec:GWs}. The summary and prospects are presented in \S\ref{sec:summary}.

\section{Fast radio bursts in a nutshell}\label{sec:FRBs}

The field of FRBs has been extensively reviewed several times \cite{katz18,popov18,platts19,petroff19,petroff22,cordes19,zhang20b,xiao21,lyubarsky21,bailes22,zhang23}. In this section, I briefly summarize the observational facts and theoretical models of FRBs and refer to the readers to \cite{zhang23} for a much more in-depth discussion of the physics of FRBs. 

\subsection{Observational facts}

The main observational properties of FRBs are summarized below. For the details of FRB observational properties, see \cite{petroff19,petroff22,cordes19,zhang23} and references therein. 

\begin{figure}[th]
\includegraphics[width=5in]{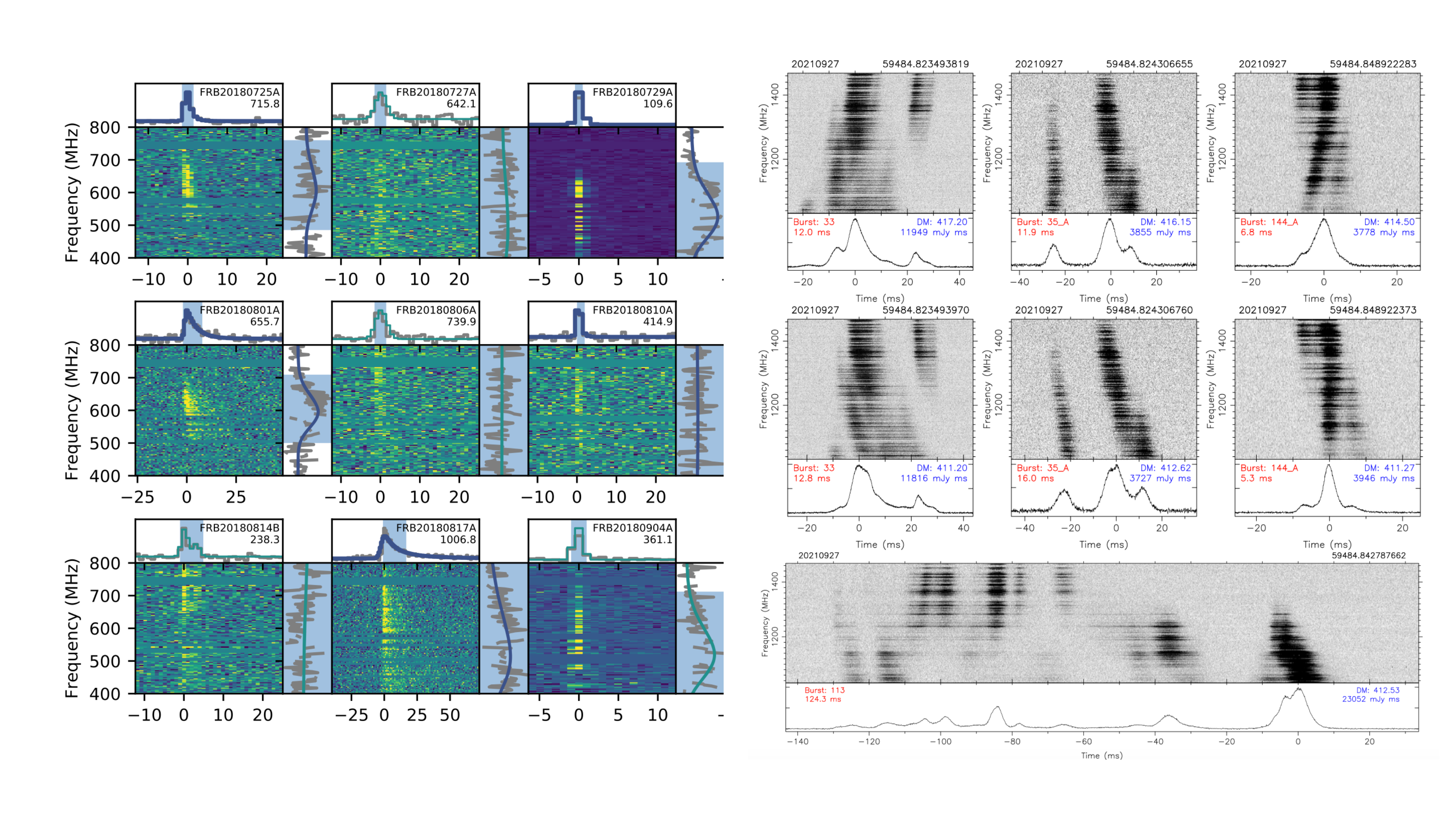}
\caption{A gallery of FRB lightcurves and their frequency-time dynamic spectra, including (a) examples of some CHIME-detected one-off FRBs \cite{chime-1st-catalog} and (b) examples of some FAST-detected repeating bursts from rFRB 20201124A \cite{zhou22}. Most one-off FRBs show single pulses; FRB 20180801A and FRB 20180817A each show a clear long tail due to scattering. Repeating bursts typically show complicated temporal and spectral properties, usually with narrower spectra than one-off FRBs. Abbreviations: CHIME: Canadian Hydrogen Intensity Mapping Experiment; DM: dispersion measure; FAST: Five-hundred-meter Aperture Spherical Telescope; FRB: fast radio burst.}
\label{fig:FRBs}
\end{figure}

\begin{itemize}
    \item The typical duration of an FRB is of the order of milliseconds. Some bursts have one single pulse \cite{lorimer07}, but bursts with multiple sub-pulses have been commonly observed \cite{champion16,pleunis21b,zhou22}. Some FRB pulses have an extended decaying tail consistent with scattering in a cold plasma \cite{thornton13}. In some bursts, frequency down-drifting of sub-pulses (also called the sad-trombone effect) have been observed \cite{hessels19,chime-repeaters,zhou22}. See Figure \ref{fig:FRBs} for a gallery of FRBs. 
    \item Even though the majority of FRB sources were detected with one single burst, a good fraction of sources show repeated bursts \cite{chime-repeaters,chime-new-repeaters}. It is possible that all FRB sources repeat and those apparent non-repeaters are just repeaters with long waiting times. However, some differences in the observational properties between repeating and non-repeating bursts have been noticed \cite{chime-repeaters}. For example, repeating bursts tend to have broader widths, narrower spectra, complicated pulse shapes and frequency down-drifting of sub-pulses. Most apparent non-repeaters, on the other hand, tend to have narrow, simple pulse profiles. 
    \item Searches for periodicity for repeater bursts usually fail \cite{zhangy18,xuh22,niujr22}. However, there were some exceptions. First, one special $\sim 3$-s long FRB 20191221A showed a 0.2168-s periodicity with 6.5$\sigma$ \cite{chime-period}. This is the only case of a periodicity that is consistent with the spin period distribution of known pulsars. Second, much longer periods (or emission cycles) have been observed in two repeating FRBs. The first is a repeater source rFRB 20180916B\footnote{Repeating FRBs are named in the format of rFRBYYYYMMDD following the convention of \cite{zhang23}.} that shows a distinct 16-d periodicity with a frequency-dependent, $\sim 5$-d duration active window \cite{chime-periodic}. Another less significant case is rFRB 20121102A, which showed a $\sim$ 160-d period \cite{rajwade20}. 
    \item FRB spectra can be fitted by either a power law function (usually for some non-repeating bursts) or a Gaussian function (usually for repeating bursts). The spectra are typically narrow for the majority of repeating bursts \cite{chime-repeaters,zhou22,zhangyk23}.
    \item FRB pulses are typically strongly polarized with nearly 100\% linear polarization \cite{michilli18,xuh22,jiangjc22}. Circular polarization has been detected in a small fraction of FRBs \cite{kumar22,xuh22,jiangjc22}. The linear polarization angle (PA) remains roughly constant in the majority of bursts \cite{michilli18,jiangjc22}, but significant PA variations across single bursts have been observed in both non-repeating and repeating bursts \cite{day20,luo20b}.
    \item The FRB dispersion measure (DM = $\int n_e/(1+z) dl$) is the column number density of free electrons along the line of sight and is therefore a good proxy of the source redshift $z$. A theoretically motivated ${\rm DM}-z$ correlation has been observationally established \cite{macquart20} after deducting the milky-way contribution to DM. 
    \item For well-localized FRB sources whose host galaxies are identified, the properties of the hosts show diverse features (from dwarf star-forming galaxies to massive MW-like galaxies) \cite{tendulkar17,bannister19,ravi19,bhandari20,lizhang20}. The positions of FRBs within their hosts also can be diverse: for instance, they may be in the star-forming region, in the off-spiral-arm region, on the outskirt, or outside the host galaxy \cite{xuh22,dong23}. In particular, one repeating source, rFRB 20200120E was discovered to be associated with a globular cluster in the nearby galaxy M81 at a distance of 3.6 kpc \cite{bhardwaj21,kirsten22}.
    \item The FRB Faraday rotation measure (RM $\propto \int B_\parallel n_e/(1+z)^2 dl $) is the line integral of parallel magnetic field and free electron number density along the line of sight. Observationally, the RMs of repeating FRBs are found to show both long-term \cite{michilli18} and short-term \cite{xuh22} variations, sometimes even sign reversals \cite{anna-thomas23}. This suggests a dynamically evolving, highly magnetized ambient environment near these FRB sources. 
    \item The observed FRB all-sky event rate and the intrinsic event rate density are both high. The all-sky rate is of the order of $10^3-10^4$ per day with the current survey sensitivity \cite{thornton13}. The event rate density can be as high as $10^7 \ {\rm Gpc^{-3} \ yr^{-1}}$ above $10^{38} \ {\rm erg \ s^{-1}}$ \cite{lu20}, exceeding that of core-collapse supernovae ($\sim 10^5 \ {\rm Gpc^{-3} \ yr^{-1}}$) by orders of magnitude, suggesting that the majority of FRBs should not be produced by cataclysmic events \cite{ravi19b,luo20}. 
    \item The global luminosity (energy) function for the FRB population is consistent with a power-law with a possible exponential cutoff at the high end \cite{luo18,luo20,lu18,lu20}. The isotropic radio energy of the observed FRBs ranges from $10^{35}-10^{43}$ erg. The redshift distribution of FRBs is still not very well constrained \cite{shin23}, but the possibility that a fraction of FRBs have a delay with respect to the star formation history of the universe has been suggested \cite{zhangzhang22,hashimoto22,qiang22}. 
\end{itemize}

\subsection{Theoretical models}

Recent comprehensive reviews on FRB theories can be found in  \cite{zhang20b,xiao21,lyubarsky21,zhang23}. Here I mainly highlight the key points following \cite{zhang23}.

\begin{itemize}
    \item All FRB models should include two ingredients: a source model that can account for the general properties of FRB population, including event rate density, redshift distribution, luminosity function, etc.; and a coherent radiation model that can account for the observational properties of individual bursts, including luminosity, duration, spectra, frequency down-drifting of sub-pulses, polarization properties, and especially the enormously high brightness temperature.  
    \item The source models include repeating models and cataclysmic models. The most popular repeating model invokes a highly magnetized neutron star, or ``magnetar'', that powers repeating bursts by drawing on the magnetic energy reservoir of the neutron star \cite{popov10,kulkarni14,katz16,lyubarsky14,metzger17,kumar17,beloborodov17,yangzhang18,metzger19,wadiasingh19,beloborodov20,lu20,margalit20,lyutikov21,yangzhang21,zhang22}. Other neutron star models include young pulsar giant-pulse models \cite{cordes16,connor16} and interacting neutron star models \cite{zhang17,dai16}, which utilize other energy sources, including the spin energy, kinetic energy, or gravitational energy to power FRBs. Beyond neutron star models, models invoking accreting black holes \cite{katz17b,sridhar21} have been discussed in the literature along with a list of more exotic models that invoke experimentally unproven physics. Furthermore, a cohort of cataclysmic models have been proposed over the years with two widely discussed groups: one invoking various types of compact binary coalescences (CBCs) \cite{totani13,mingarelli15,zhang16a,wang16,levin18,zhang19,dai19}; and the other invoking implosions of supermassive neutron stars (SMNSs) that eject the magnetopsheres of the stars (also termed as ``blitzars'') \cite{falcke14,zhang14,most18}.
    \item The detection of FRB 20200428D from the Galactic magnetar SGR J1935+2154 suggests that at least some FRBs are made by magnetars. The question is whether {\em all} FRBs are made by magnetars. One may consider two extreme views \cite{zhang20b}. The most conservative view is that magnetars can make them all, regardless of whether they are repeating or non-repeating. Since there is no active repeaters observed from the Milky Way, one needs to invoke some special magnetars not observed from the Milky Way to account for active repeaters from other galaxies. One likely candidate is the new-born, rapidly spinning magnetars resulting from extreme explosions such as gamma-ray bursts, superluminous supernovae, and binary neutron star mergers. On the other hand, a more speculative view suggests that there are multiple channels to power FRBs from both repeating and intrinsically non-repeating (cataclysmic) sources. 
    \item Growing observations posed challenges to the most conservative view that ``magnetars make them all'': Since there are no active repeaters in the Milky Way galaxy and since none of the Milky Way magnetars are in globular clusters, one needs to have at least three types of magnetars to produce the observed FRB population: besides the regular magnetars that can produce FRB 20200428D-like events, one needs to have some special magnetars (younger or in more special environments than Galactic magnetars) to account for the active repeater population; and one also needs to have a new channel to form magnetars in the old stellar population (e.g. WD-WD mergers \cite{kremer21,lu22,kremer23}). 
    Some active repeaters reside in dynamically evolving, magnetized environments that are more complicated than a supernova remnant or magnetar wind nebula \cite{xuh22}. Some other active repeaters, on the other hand, seem to reside in a plain and not highly magnetized environment \cite{zhangyk23}. Also, while rFRB 20180916B has a clear 16-d period, most other active repeaters do not show a clear periodicity. All these add puzzles to the simple magnetar picture. 
    \item Within the popular magnetar source model, multiple scenarios have been suggested in the literature to power repeating FRBs. Based on the proposed emission region of FRBs, the models can be categorized into closer-in (or pulsar-like) models in which FRBs are produced inside or slightly outside the magnetar magnetosphere \cite{kumar17,yangzhang18,wadiasingh19,kumar20a,lu20,lyubarsky20,yangzhang21,lyutikov21,zhang22,qu22b,qu23}, and farther-out (or GRB-like) models in which FRBs are produced in relativistic shocks far away from the light cylinder of the magnetar \cite{lyubarsky14,beloborodov17,metzger19,beloborodov20,sironi21}. The magnetospheric models include bunched coherent emission due to curvature radiation \cite{kumar17,yangzhang18,wangwy22,wangwy22b}, inverse Compton scattering \cite{zhang22,qu23,quzhang23}, free electron laser \cite{lyutikov21}, and plasma instabilities inside the magnetosphere \cite{lu18} or in the current sheet region \cite{lyubarsky20} outside the magnetosphere. Growing observational data (e.g. polarization angle swing \cite{luo20b} and significant circular polarization \cite{xuh22}, short waiting time for active repeaters \cite{lid21}, large radio burst energetics demanding a high radio emission efficiency \cite{zhangyk22,zhangyk23}) seem to support the magnetospheric origin of most bursts. There was a theoretical objection to the magnetospheric model stating that bright FRBs cannot escape the magnetosphere because of the huge scattering optical depth \cite{beloborodov21,beloborodov23}. A closer investigation of this effect suggests that FRBs can escape the magnetosphere if they are produced in the open field line region where the background plasma is flowing out of the magnetosphere with a relativistic speed \cite{qu22b,lyutikov23}. Observationally, only a very small fraction of magnetar X-ray bursts are associated with FRBs \cite{lin20}. This suggests that the physical condition that allows FRBs to form and propagate is quite demanding. Modelers are allowed and even encouraged to introduce special physical conditions to produce FRBs from a magnetar. 
    \item Among the non-magnetar models, those of particular interest to this review are the models that invoke associations between FRBs or FRB-like objects and gravitational waves, mostly related to binary neutron star (BNS, i.e. NS-NS) mergers, but also neutron star - black hole (NS-BH) and black hole - black hole (BH-BH) mergers as well. These models distinguish themselves from other models to have falsifiable predictions of multi-messenger signals, and are physically well motivated. There are both repeating and cataclysmic models. For repeating models, besides the straightforward BNS-merger-born young magnetar models \citep{margalit19,wang20}, it was suggested that magnetospheric interactions between two NSs decades to centuries before their merger may be also able to power repeating FRBs \cite{zhang20}. For non-repeating models, FRBs are proposed to be generated during the magnetosphere synchronization during the merger \cite{totani13} or when a post-merger SMNS collapses to a black hole minutes to hours after the merger \cite{zhang14}. An array of models invoke a rapidly increasing Poynting flux from a CBC system right before the merger, which includes the unipolar effect during the final stage of magnetospheric interactions of BNS mergers \cite{piro12,lai12,wang16}, charged CBC effect for any merging systems with any member being charged \cite{zhang16a,liebling16,deng18,zhang19}, black hole battery effect \cite{levin18,dai19}, as well as merger-induced charge discharge \cite{liu16} or reconnection \cite{fraschetti18}. These models will be discussed in detail in \S\ref{sec:GWs}.
\end{itemize}

\section{Multi-wavelength emissions from FRBs}\label{sec:MW}

I now discuss the multi-wavelength counterparts of FRBs. Because the origin of FRBs is still up in the air, there is a wide range of predictions involving different energy bands and different time windows. I group these predictions in three main categories: prompt emission signals that temporarily coincide with FRBs or somewhat offset from FRBs, afterglows that follow FRBs, counterparts associated with the birth of the FRB engines, and persistent counterparts. In the following we discuss these four types of counterparts in turn. 

\subsection{Prompt emission counterparts}

\begin{figure}[th]
\includegraphics[width=5in]{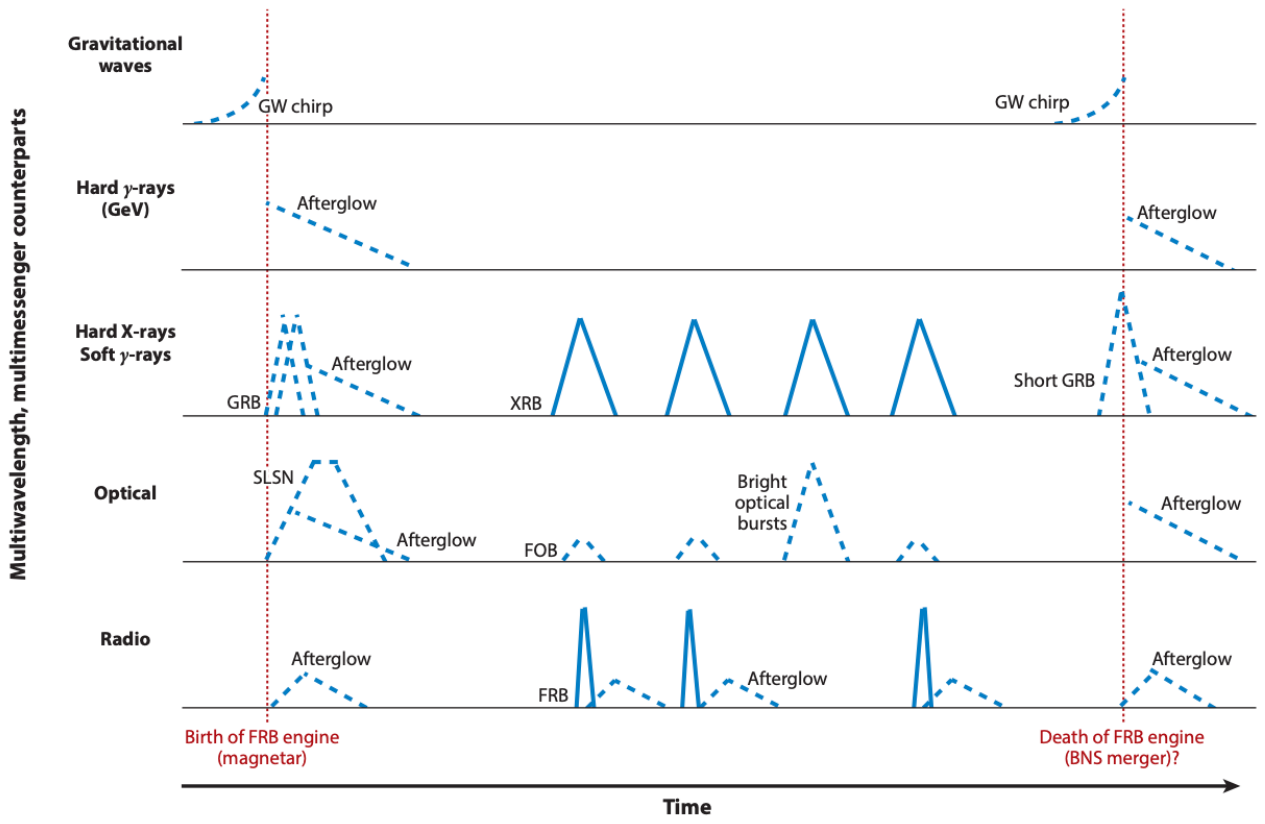}
\caption{Various multi-wavelength counterparts of FRBs. Solid lines denote detected events whereas dashed lines denote speculated counterparts. Persistent counterparts (PRSs) are not shown. Abbreviations: BNS: binary neutron star; FOB: fast optical burst; FRB: fast radio burst; GRB: $\gamma$-ray burst; GW: gravitational wave; PRS: persistent radio source; SLSN: superluminous supernovae; XRB: X-ray burst.}
\label{fig:multi-wavelength}
\end{figure}

The brightness temperature, $T_b$, of a radio source is the temperature the source would have if the emission spectrum were a blackbody. For non-thermal emission emission, $T_b$ can be much greater than any known true temperatures of astronomical bodies, and it tell how much the emitting particles deviate from thermal equilibrium. The $T_b$ of FRB emission can be estimated as from its flux, distance, and duration (which carries the information of the size of the source), which reads \cite{luojw23,zhang23}
    \begin{eqnarray}
        T_b & \simeq & \frac{{\cal S}_{\nu,p} D_{\rm A}^2(1+z)^3}{2 \pi k_B (\nu \Delta t)^2} = (1.2\times 10^{36}  \ {\rm K} ) (1+z)^3
        %\nonumber \\
        \left( \frac{D_{\rm A}}{10^{28} \ {\rm cm}} \right)^2 \frac{{\cal S}_{\nu,p}}{\rm Jy} \left(\frac{ \nu}{\rm GHz}\right)^{-2} \left(\frac{\Delta t}{\rm ms}\right)^{-2},
        \label{eq:Tb}
    \end{eqnarray}
where $k_B$ is Boltzmann constant, $S_{\nu,p}$ is the peak specific flux in units of Jansky (1 Jy = $10^{-23} \ {\rm erg \ s^{-1} \ Hz^{-1}}$), $D_{\rm A}$ is the angular distance of the source, $z$ is redshift, $\nu$ is the characteristic observing frequency, and $\Delta t$ is the duration of the burst. This is much higher than the maximum $T_b$ allowed for incoherent radiation, which is $T_{\rm b,max}^{\rm incoh}
=(5.9\times 10^{13} \ {\rm K}) \gamma_{e,2} \Gamma_2$, where $\gamma_e$ is the characteristic electron Lorentz factor and $\Gamma$ is the bulk Lorentz factor, both normalized to 100. As a result, FRB emission 
requires non-thermal particles to radiate coherently. Even though a variety of coherent mechanisms have been discussed in the literature without a consensus in the community, a common feature of all these models invoke relativistic particles (either the same population that produce FRBs themselves or a different associated population) that can radiate at other frequencies, typically at much higher frequencies. Such high-frequency emission components are discussed below (see Figure \ref{fig:multi-wavelength} for various multi-wavelength counterparts discussed in the literature). 

\subsubsection{Hard X-rays/soft $\gamma$-rays -- FRB 20200428D and FRB/XRB association}\label{sec:Xrays}

We first discuss hard X-rays/soft $\gamma$-rays because this is the only energy band from which an FRB counterpart was detected, from the Galactic FRB 20200428D.

The motivation to suspect a hard X-ray/soft $\gamma$-ray component to an FRB was based on the belief that magnetars can produce FRBs. The first suggestion on the connection \cite{popov10} was based on the observation that magnetars occasionally make ``giant flares'', which typically consist of a short-hard spike and an extended oscillating tail. The short-hard soft $\gamma$-ray spike typically has a duration of milliseconds, consistent with that of an FRB. As a result, it is straightforward to speculate that FRBs are produced in association with the short-hard spikes of magnetar giant flares. Such a suggestion was put in doubt by the non-detection of radio emission during the 2004 December 27 giant flare of SGR 1806-20 \cite{tendulkar16}, with a tight fluence upper limit inconsistent with the fluences of known FRBs at cosmological distances.  As a result, for some time, the magnetar model did not appear particularly appealing and was regarded only as one of many possibilities. Before the detection of FRB 20200428D, some researchers (myself included) did not believe that regular magnetars can make FRBs. Indeed, several days before the occurrence of FRB 20200428D, the Five-hundred-meter Aperture Spherical Telescope (FAST) in China had been monitoring the magnetar source SGR J1935+2154 for a few days after it became $\gamma$-ray active. The expectation was to place a more stringent radio upper limit during one the bursts.  It turned out to be the case: very stringent radio pulse flux limits were placed during 29 hard X-ray bursts during one of the FAST observations \cite{lin20}. Therefore, the detection of FRB 20200428D \cite{CHIME-SGR,STARE2-SGR} in association with another hard X-ray/soft $\gamma$-ray burst \cite{HXMT-SGR,Integral-SGR,Konus-SGR,AGILE-SGR} turned out to be a surprise for several reasons: 1. The associated hard X-ray burst (XRB) was not a giant flare but rather a regular burst;  2. Among hard XRBs emitted from magnetars, this burst was not the brightest one \cite{yangyh21}; 3. The duration of the XRB was much longer than milliseconds. It is the two narrow peaks of the XRB that correspond to the two pulses of FRB 20200428D detected by CHIME (with slight time offsets). Nonetheless, a closer scrutiny at the XRB revealed a relatively hard spectral peak energy \cite{younes21} and a possible quasi-periodic-oscillation signature \cite{lixb22} in the emission of the burst, which suggests that the XRB may be special. Alternatively, the lack of FRB detection in other XRBs may be interpreted as being due to a much narrower beaming angle of FRBs than XRBs \cite{lin20}. Within such an interpretation, off-axis FRBs with longer durations and softer spectra (termed as ``slow radio bursts'' or SRBs \cite{zhang21}) are expected from SGR J1935+2154. A search for such events led to a constraint of $\theta_j \Gamma \sim $ a few, where $\theta_j$ is the opening angle of the FRB jet and $\Gamma$ is its bulk Lorentz factor \cite{chenzhang22}. Modeling FRB propagation within magnetar magnetospheres attributes the lack of FRB association with the giant flare of SGR 1806-20 to choked FRB emission by a bright fireball associated with the giant flare \cite{ioka20b}. According to this calculation, FRBs are preferably associated with not-too-bright XRBs. 

Since there is only one case of FRB-XRB association, it is informative to summarize the observational properties of this association and set it as a standard for searching for other possible associations.

\begin{itemize}
    \item FRB 20200428D as detected by CHIME in the 400-800 MHz band \cite{CHIME-SGR} has two sub-bursts with durations of $0.585 \pm 0.014$ ms and $0.335 \pm 0.007$ ms, respectively, separated by $28.91 \pm 0.02$ ms. The peak flux densities of the two sub-bursts are 110 kJy and 150 kJy, respectively, and the fluences are $480 {\rm kJy \ ms}$ and $220 {\rm kJy \ ms}$, respectively. The distance of the source is not well measured, which ranges from 6.6 to 12.5 kpc. With a fiducial 10 kpc distance and in the 400-800-MHz band, the burst has a total isotropic energy of $(3^{+3}_{-1.6} \times 10^{34} \ {\rm erg}) \ d_{\rm 10 kpc}^2$ and a peak isotropic luminosity of $(7^{+7}_{-4} \times 10^{36} \ {\rm erg \ s^{-1}}) \ d_{\rm 10 kpc}^2$.
    \item FRB 20200428D as detected by STARE2 in the 1.281-1.468 GHz band \cite{STARE2-SGR} has only one peak with a fluence $(1.5 \pm 0.3) {\rm MJy \ ms}$, corresponding to an isotropic energy of $(2.4 \pm 0.4) \times 10^{35} \ {\rm erg}$ for a fiducial distance of 10 kpc. The burst has an intrinsic width of $0.61\pm 0.09$ ms. This would bring the isotropic radio average luminosity to $\sim (4 \times 10^{38} \ {\rm erg \ s^{-1}}) d_{\rm 10 kpc}^2$. The arrival time of this burst coincides with the second sub-burst as detected by CHIME. Since the spectral shape of that sub-burst is rising in the CHIME band, it is fully consistent with the second sub-burst of CHIME being the same burst detected by STARE-2. This energy/luminosity is lower by a factor of a few tens than the faintest cosmological FRBs, but the rate of occurrence is fully consistent with extending the FRB burst rate to lower energies/luminosities \cite{STARE2-SGR,lu20}.
    \item The XRB associated with FRB 20200428D was simultaneously detected by a few X-ray missions. The 1-250 keV unabsorbed fluence as detected by Insight-HXMT \cite{HXMT-SGR} is $(7.14^{+0.41}_{-0.38} \times 10^{-7}) \ {\rm erg \ cm^{-2}}$, corresponding to $\sim (0.85\times 10^{40} \ {\rm erg}) \ d_{\rm 10 kpc}^2$ isotropic energy for a fiducial distance $d=10$ kpc. The total duration is 0.53 s, much longer than the duration of the FRB. There are two hard spikes in the lightcurve with a separation of $\sim 31.5$ ms, whose arrival times are broadly consistent, but slightly delayed \cite{ge23}, with respect to the two CHIME FRB sub-bursts (with a separation of $\sim 29$ ms) after dispersion correction. Those two peaks can be taken as the counterparts of the FRBs.  Other X-ray mission teams reported similar results: In the 20-200 keV INTEGRAL band \cite{Integral-SGR}, the fluence was $(6.1\pm 0.3)\times 10^{-7} \ {\rm erg \ cm^{-2}}$ and the duration was $\sim 0.6$ s. The two CHIME sub-bursts again broadly track two X-ray peaks (with the radio sub-bursts leading by $6.5\pm 1.0$ ms). Konus-Wind observations \cite{Konus-SGR} gave a 20-500 keV fluence $(9.7\pm 1.1) \times 10^{-7} \ {\rm erg \ cm^{-2}}$ and a peak flux (second peak) $(9.1\pm 1.5) \times 10^{-6} \ {\rm erg \ cm^{-2} \ s^{-1}}$, which correspond to a total isotropic X-ray energy of $(1.2 \times 10^{40} \ {\rm erg}) d_{\rm 10 kpc}^2$ and a peak isotropic energy luminosity $(1.1 \times 10^{41} \ {\rm erg \ s^{-1}}) d_{\rm 10 kpc}^2$. Finally, AGILE observations \cite{AGILE-SGR} gave an 18-60 keV fluence of $\sim 5 \times 10^{-7} \ {\rm erg \ cm^{-2}}$, an isotropic energy $\sim (0.81 \times 10^{40} \ {\rm erg}) d_{\rm 10 kpc}^2$, and a duration shorter than 50 s. 
    \item Combining X-ray and radio data, in particular the Konus-Wind and STARE-2 results (which respectively report the largest values), one gets the reference values of the energy and luminosity ratios between the two bands for this pair of association:
    \begin{eqnarray}\label{eq:standard}
     \eta_{\rm X/r}^{\rm E}  \equiv \frac{E_{\rm X}}{E_r} 
     %\sim (5\times 10^{-13}) \frac{\rm erg \ cm^{-2}}{\rm Jy \ ms \ GHz} 
     \sim 5\times 10^4, ~~~ 
     %\nonumber \\
     \eta_{\rm r/X}^{\rm E}  \equiv \frac{E_r}{E_{\rm X}}
     %\sim (2\times 10^{12}) \frac{\rm Jy \ ms \ GHz}{\rm erg \ cm^{-2}} 
     \sim 2\times 10^{-5}, \nonumber \\
     \eta_{\rm X/r}^{\rm L}  \equiv \frac{L_{\rm X}}{L_r} 
     %\sim (3\times 10^{-12}) \frac{\rm erg \ cm^{-2} \ s^{-1}}{\rm Jy \ GHz} 
     \sim 3\times 10^2,  
     %\nonumber \\
     ~~~ 
     \eta_{\rm r/X}^{\rm L}  \equiv \frac{L_r}{L_{\rm X}} %\sim (3\times 10^{11}) \frac{\rm Jy \ GHz} {\rm erg \ cm^{-2} \ s^{-1}} 
     \sim 3\times 10^{-3}
     %\nonumber \\
     %~~~ {\rm for~FRB~20200428}
    \end{eqnarray}
for FRB 20200428D, noticing $\rm (erg \ cm^{-2}) / (Jy \ ms \ GHz) = 10^{17}$ and $\rm (erg \ cm^{-2} \ s^{-1}) / (Jy \ GHz) = 10^{14}$. Note that $\eta^{\rm E}_{\rm X/r}$ is more than two orders of magnitude larger than $\eta^{\rm L}_{\rm X/r}$, which can be accounted for noticing that the XRB duration is more than two orders of magnitude longer than the total duration of the two FRB sub-pulses.  Hereafter, we refer to the relations in Eq.(\ref{eq:standard}) as the ``FRB 20200428D standard'' for X-ray emission and use them as guidance for other possible associations as discussed below.
\end{itemize}

\subsubsection{X-rays/$\gamma$-rays -- other cases}\label{sec:Xrays2}

\begin{figure}[th]
\includegraphics[width=5in]{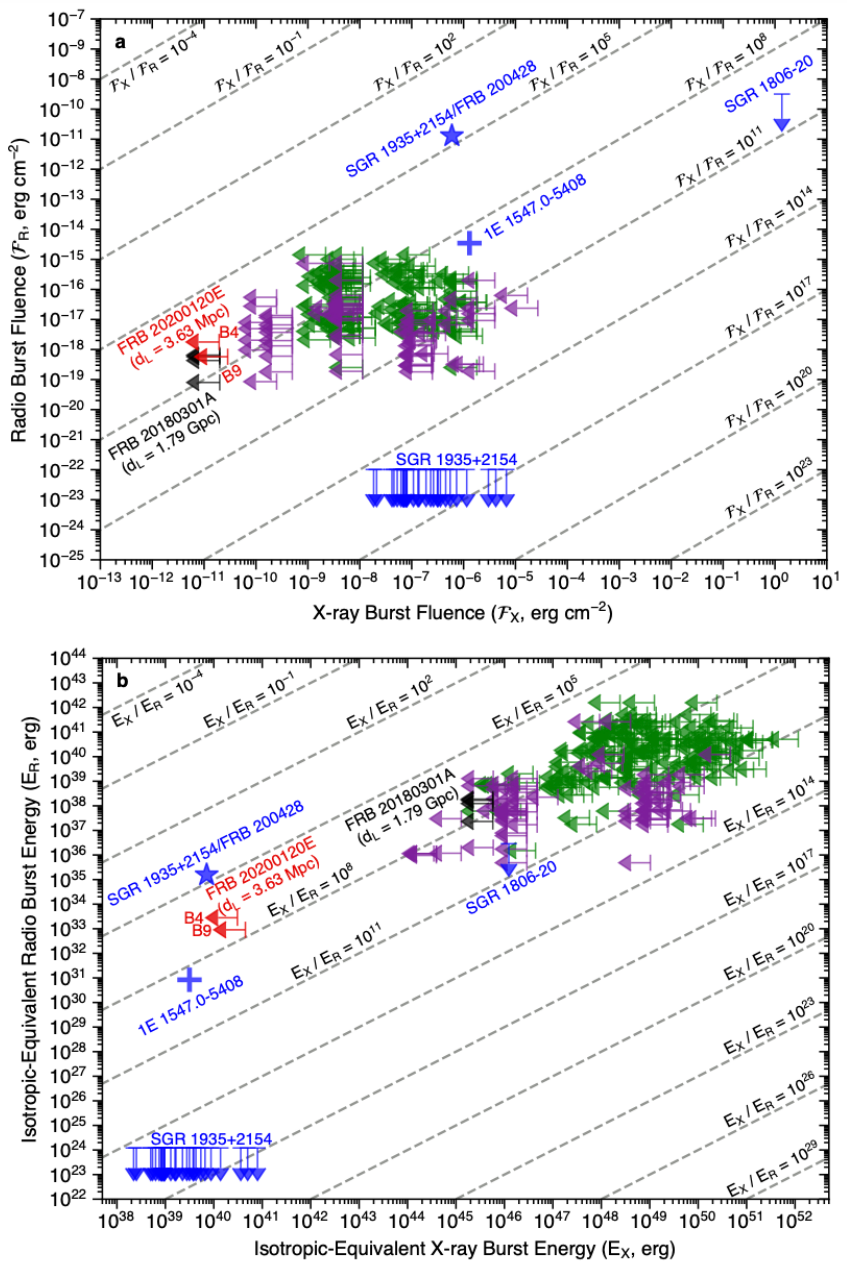}
\caption{Various X-ray/$\gamma$-ray fluence/energy upper limits for cosmological FRBs compared with the detection of FRB 20200428D from the Galactic magnetar SGR J1935+2154. Figure adapted from Ref.~\cite{pearlman23}.}
\label{fig:X-rays}
\end{figure}

For cosmological FRBs, simultaneous X-ray/soft $\gamma$-ray observations using detectors such as Swift/XRT and NICER have been carried out, which so far only led to upper limits \cite{scholz17,scholz20,laha22a,laha22b,piro21,pearlman23}. One immediate inference from the FRB 20200428D standard (Eq.(\ref{eq:standard})) is that a bright Jy ms level FRB at a cosmological distance would correspond to an X-ray fluence of $5\times 10^{-13} \ {\rm erg \ cm^{-2}}$, which is several orders of magnitude below all the triggering $\gamma$-ray detectors (Swift BAT, Fermi GBM, GECAM, Insight-HXMT, Integral, Konus-Wind, AGILE, EP WXT, SVOM ECLAIRS, etc). The only way of detecting the X-ray component of these FRBs is to utilize more sensitive imaging telescopes such as Swift XRT, EP FXT, NICER, Chandra, XMM-Newton, etc. This requires coordinated joint observations between radio and X-ray telescopes with the hope that some radio bursts will occur during X-ray observations. Even with such observations, usually the FRB 20200428D standard is hard to beat for individual bursts. Stacked searches of X-ray emission have also been performed. Yet, no detection is made so far. The latest status is well summarized in \cite{pearlman23} (see Fig.\ref{fig:X-rays}). The fact that none of the current upper limits have beaten the FRB 20200428D standard suggests that the current non-detection is consistent with the hypothesis that all FRBs have the same or lower X-ray-to-radio fluence (or energy) ratio. In terms of absolute X-ray burst energy, the M81 globular cluster upper limit is the tightest, but is still above that of FRB 20200428D. 

More generally, searches for GRB-like emission around the time of FRBs have been made, mainly motivated by some tentative but not confirmed early possible associations \cite{bannister12,delaunay16}.
Later searches all led to non-detection \cite{cunningham19,martone19,anumarlapudi20,casentini20,guidorzi20,tohuvavohu20,mereghetti21,verrecchia21,curtin23}. Based on an energetics argument, i.e. FRBs are much less energetic and far more frequent than catastrophic events, it is not expected that the majority of FRBs could be associated with GRB-like emission. The non-detection is consistent with such an expectation. It also places some constraints on some models that predict GRB-FRB associations in a small fraction of FRBs. See \cite{curtin23} for detailed discussion. 

At even higher energies (GeV $\gamma$-rays), upper limits have been placed on some repeating FRBs based on the Fermi LAT data, posing lose constraints on the FRB and FRB progenitor models \cite{zhangzhang17,yangyh19}. See \S\ref{sec:foreglow} for more discussion.

\subsubsection{Optical emission}\label{sec:opt} 

Prompt FRB counterpart in the optical band has been discussed within the framework of some theoretical models. The most straightforward counterpart is the one that temporarily associated with an FRB itself. Such a ``fast optical burst'' could be the spectral extrapolation of FRB emission to the optical band or the inverse Compton scattering component of the FRB emission to higher frequencies \cite{yang19}. The simple extrapolation is not physically motivated because the radio emission is strongly coherent and is likely a distinct component. The extrapolated flux is also very low. A better chance of detecting such prompt optical emission is through inverse Compton scattering of radio photons by electrons internal (one-zone) or external (two-zone) to the FRB emission region. In any case, the prospect of detecting prompt optical emission associated with FRBs is quite low \cite{yang19}.

Within some FRB models, on the other hand, relatively bright optical emission may come from repeating FRB sources, not necessarily temporarily associated with FRB bursts. One scenario to produce bright optical flares is within the framework of the synchrotron maser model \cite{beloborodov20}. According to this model, an active magnetar may sporadically ejecta highly magnetized blobs (or ``blasts'' as called in \cite{beloborodov20}). If a blast collides with a  cold magnetar wind, the reverse shock emission in the blastwave would generate a radio burst through the synchrotron maser mechanism. On the other hand, if two blasts have a short separation and the late blast impacts on the tail of a previous flare, the emission would be in the optical band because the upstream is hot. The luminosity of such an optical flare could be as bright as $10^{44} \ {\rm erg \ s^{-1}}$.  Another scenario \cite{yangyp21} to produce moderately bright optical flares is within the binary model for FRB sources, as has been suggested to account for either the apparent 16-d period of rFRB 20180916B \cite{ioka20,lyutikov20} or the short-term RM variations \cite{wangfy22,anna-thomas23}. According to this scenario, if FRB jets are highly beamed and can point toward random directions (as suggested by the data \cite{zhu23}), a very small fraction of FRB jets may hit the companion star, making a bright optical flare upon depositing a huge amount of energy to the star \cite{yangyp21}. Such optical flares can be brighter than the Sun by several times and last for hundreds of seconds. 

Observationally, many optical upper limits simultaneous to FRB emission have been posed \cite{hardy17,MAGIC18,lin20,andreoni20,niino22,marcote20,kilpatrick21,hiramatsu23}. The tightest upper limit is the one set to FRB 20200428D from the Galactic magnetar SGR J1935+2154 \cite{lin20}. This was a $Z$-equivalent 17.9 magnitude upper limit in a 60-s exposure during the prompt epoch of FRB 20200428D obtained with the BOOTES-3 telescope. After extinction correction, this upper limit is $\sim 11.7$ mag, which is effectively $\sim 4.4$ kJy if one assumes a 1-ms timescale for optical emission. This gives the ``FRB 20200428D standard'' for optical emission (for 1 ms FOB emission) \cite{lin20}
\begin{eqnarray}
    \eta_{\rm O/r}^{\rm L} & \equiv & \frac{L_{\rm \nu,O}}{L_{\rm nu,r}} = \frac{F_{\rm \nu,O}}{F_{\rm nu,r}} < 10^{-3}, ~~~
    {\rm or} \nonumber \\ 
    L_{\rm O} & \simeq & (\nu L_\nu)_{\rm O} < (7\times 10^{33} {\rm erg \ s^{-1}}) \left( \frac{\nu_{\rm O}}{\nu_{\rm r}} \right) \simeq  3\times 10^{39} \ {\rm erg \ s^{-1}}.
\end{eqnarray}
Compared with Eq.(\ref{eq:standard}), one can see that FRBs are much poorer emitters in the optical band than in the X-ray band. This upper limit is consistent with most prompt FOB models discussed in \cite{yang19}, but starts to constrain the parameter space of the inverse Compton scattering model \cite{lin20}. The bright optical flares predicted in the synchrotron maser shock model \cite{beloborodov20} is ruled out in this case. 

For cosmological FRBs, because of their large distances, the luminosity upper limits are much looser thanks to their large distances, mostly in the range of $10^{43} - 10^{45} \ {\rm erg \ s^{-1}}$. One of the tightest constraints was achieved in rFRB 20220912A \cite{hiramatsu23} with $L_{\rm O} < 10^{42} (2\times 10^{41} \ {\rm erg \ s^{-1}}$) for millisecond to second timescales. Due to the high luminosity of FRBs, the relative ratio $\eta_{\rm O/r}^{\rm L}$ can reach 0.02, about one order of magnitude higher than the optical FRB 20200428D standard for 1-ms FOB emission \cite{hiramatsu23}. Much looser physical constraints are posed on cosmological FOB models \cite{yang19}, and the synchrotron maser model \cite{beloborodov20} is again disfavored by some of the tight constraints. 

\subsubsection{Emission above 10 GHz and below 100 MHz}\label{sec:radio}

So far, FRBs are detected between 110 MHz \cite{pleunis21} and 8 GHz \cite{gajjar18}. The non-detection of FRBs above 10 GHz or below 100 MHz may be related to instrumental selection effects, but there could be intrinsic deficit of these sources as well (e.g. difficulty to produce coherent emission at higher frequencies and stronger absorption processes at lower frequencies). It is now clear that repeating FRBs tend to have narrow spectra \cite{chime-repeaters,zhou22,zhangyk23}, so  the same burst may not be detected by different telescopes with different frequency ranges. Detecting bursts beyond the current boundaries would tell us more about the extreme physical and environmental conditions to produce FRBs. 

\subsection{Afterglows} 

GRBs are typically followed by multi-wavelength afterglow emission. This is based on a generic argument that a relativistic fireball should be decelerated by the ambient medium driving a shock from which electrons are accelerated and emit synchrotron radiation \cite{meszarosrees97,sari98}. Various arguments suggest that FRBs are generated by plasmas moving with relativistic speeds (\cite{zhang23} and references therein). However, the FRB afterglows are expected to be too faint to be detected \cite{yi14}. The maximum spectral flux of synchrotron radiation, $F_{\rm \nu,max}$, is proportional to the isotropic-equivalent kinetic energy $E_{\rm iso}$ of the fireball for a constant density medium. Observationally, a typical FRB isotropic emission energy ($10^{38}$ erg) is about 13 orders of magnitude smaller than that of a typical long GRB ($10^{52}$ erg). As a result, even if the kinetic energy of an FRB much greater, e.g. $E_{\rm iso} \sim 10^{44} \ {\rm erg}$, which is 6 orders of magnitude larger than the radio energy and 20 times greater than the FRB 20200428D X-ray standard, the predicted afterglow level is still negligibly low. A possible detection requires an extreme scenario: e.g. an energetic FRB with an extremely low radio efficiency with $E_{\rm iso}$ reaching the level of $\sim 10^{47} \ {\rm erg}$ and if there exists a highly magnetized reverse shock \cite{yi14}. Such a condition is difficult to satisfy for the majority of FRBs for the energy budget reason. 

Because of the low energy of an FRB fireball, the magnetic field strength in the shocked region is typically low, so that the afterglow emission, if any, likely peaks in the radio band. Even for the Galactic FRB 20200428D, there was no radio afterglow detected, consistent with the model prediction. For reference, a radio afterglow was detected to follow the 2004 December 27 giant flare of the Galactic magnetar SGR 1806-20 that had an isotropic $\gamma$-ray energy $>10^{46}$ erg \cite{taylor05}. The non-detection of radio afterglow from FRB 20200428D suggests that the X-ray emission (with isotropic energy $E_{\rm X} \sim 1.2 \times 10^{40}$ erg) has an efficiency much greater that $10^{-6}$. 

\subsection{Progenitor counterparts}\label{sec:foreglow}

The leading scenario to interpret repeating FRB sources invokes highly magnetized neutron stars or magnetars. This scenario has been supported by the identification of the Galactic magnetar SGR J1935+2154 as the source of FRB 20200428D \cite{CHIME-SGR,STARE2-SGR,HXMT-SGR,Integral-SGR,Konus-SGR,AGILE-SGR}. However, Galactic magnetars are typically $10^3-10^4$ years old \cite{kaspi17}. Detecting the birth events of these magnetars is out of reach for these sources. On the other hand, it has been speculated that active repeating FRB sources may originate from much younger magnetars, which are years to decades-old \cite{metzger17}. If this is the case, then one expects a catastrophic explosion, probably a supernova or a GRB, that preceded a repeating FRB source. In particular, the host galaxy of rFRB 20121102A mimics that of a long GRB or a superluminous supernova (SLSN) \cite{tendulkar17}, raising the possibility that long GRBs or SLSNe are the progenitor systems of active repeaters. It is now clear that the majority of FRBs do not have such hosts, but may have more moderate progenitor systems such as Type II supernovae. 

Efforts have been made to seek a connection between FRBs and their preceding SLSNe, GRBs, or other catastrophic events. However, no firm association has been identified. Three approaches have been implemented: 1. The first approach is to directly search for FRBs in the remnants of SLSNe \cite{law19} and GRBs that appear to have a magnetar central engine \cite{men19,madison19}, typically years after the explosions. Motivated by the suggestion that a one-off FRB may be produced shortly after some GRBs when the magnetar engine collapses to a black hole \cite{zhang14}, searches were also carried out shortly after several GRBs \cite{rowlinson19,bouwhuis20,palliyaguru21}. All these searches led to negative results. 2. The second approach is to search for spatially associated GRB-FRB or SN-FRB pairs, with the GRB or the SN leading the FRB. After extensive searches, some candidate cases have been presented in the literature. One example is the GRB 110715A - FRB 171209 pair \cite{wangxg20}. The GRB has a redshift of 0.82 and the FRB has a DM of $1457.4 \pm 0.03 \ {\rm pc \ cm^{-3}}$, which is signficiantly larger than the intergalactic medium contribution to DM at that redshift. One needs to introduce a local source, presumably a supernova remnant to account for the extra DM. Overall, the significance of the association is below $3\sigma$. Another example is the apparent association between an optical transient AT2020hur and the periodic, repeating FRB 20180916B \cite{lil22}. However, the extremely high luminosity of the optical transient defies essentially all possible physical scenarios. 3. The third approach is indirect. It is known that some active repeating FRBs are associated with a persistent radio source (PRS) \cite{chatterjee17,niu22}. If these PRSs are the remnants of some explosions, then there might be associations between FRBs with some other PRS-like radio sources in the sky. Searches for FRB emission from some of these sources have been carried out but with no postive detection so far \cite{law22}.

\subsection{Persistent counterparts}\label{sec:persistent}

Three repeating sources, rFRB 20121102A, rFRB 20190520B, rFRB 20201124A, are known to be associated with persistent radio sources (PRSs) \cite{chatterjee17,niu22,bruni23}. The origin of persistent radio emission is not identified, but it is very likely due to synchrotron emission of relativistic electrons from a surrounding nebula, possibly a supernova remnant, a pulsar wind nebula, or a hyper-accreting X-ray binary \cite{yang16,murase16,metzger17,sridhar22}. Theoretically, the brightness of a PRS is supposed to be positively related to RM of the FRB within the synchrotron nebula model \cite{yang20b,yang22}.  This prediction is consistent with the observational data so far \cite{bruni23}. 

The fact that some repeating FRBs are have PRS associations led to the hypothesis that some PRS-like radio sources in the sky might be FRB emitters. Searches for FRB emission from some of these sources have been carried out but with no positive detection so far \cite{law22}.

Another suggestion was that repeating FRB sources could be produced by hyper-accreting binaries, so that they chould be ultra-luminous X-ray sources (ULXs) \cite{sridhar22}. Deep X-ray flux upper limit from the nearby M81 globular cluster source rFRB 20200120E has ruled out this scenario at least for this repeating source  \cite{pearlman23}. 

\section{Neutrino emission from FRBs?}\label{sec:nus}

To be a high-energy neutrino emitter, an astronomical source usually needs to accelerate baryons to high energies. These baryons also need to stay in a photon or baryon target field with a high enough number density so that copious neutrinos can be emitted through $p\gamma$ or $pp$ interactions. FRBs as we know them may not be strong hadronic sources. Even if magnetars or pulsars are promising accelerators, they are mostly known to accelerate leptons in the form of electron-positron pairs. Nonetheless, it has been hypothesized that baryons may be accelerated from these sources, especially during explosions that power FRBs. The interaction between accelerated protons and background photons or baryons may produce neutrinos.

Proton acceleration and neutrino production from Galactic magnetars were first discussed in \cite{zhang03}. For the case of ${\bf \Omega \cdot B} < 0$ in the polar cap region ($\bf \Omega$ is the angular velocity vector and ${\bf B}$ is the magnetic field vector), protons can be accelerated from the surface. In order to meet the $\Delta$-resonance condition for $p\gamma$-interaction with the near-surface X-ray photons, the electric potential across the polar cap of the magnetar has to be greater than a threshold value. This defines a neutrino emission ``death line'' in the $P-\dot P$ diagram of pulsars/magnetars ($P$ is the spin period and $\dot P$ is the spindown rate), above which neutrino emission is possible. A further investigation of the flux, however, suggests that these neutrino sources are usually too faint to be detected with the available neutrino detectors (e.g. IceCube) and all magnetars in the universe only contribute to a small fraction of the neutrino background. 

The possibility of magnetar neutrino emission during the FRB phase was recently reinvestigated. Ref. \cite{metzger20} showed that within the framework of the relativistic shock synchrotron maser model, protons can be accelerated to produce neutrinos via $p\gamma$ interactions. However, the neutrino flux is many orders of magnitude below the IceCube sensitivity even for the Galactic FRB 20200428D. In general, the neutrino flux is higher if $p\gamma$ interactions happen at a smaller radius where the target photon density is higher. Ref. \cite{qu22a} investigated three scenarios of FRB generation and proton acceleration with different emission sites: inside the magnetosphere, in the current sheet region beyond the light cylinder, and in relativistic shocks. They indeed found that the predicted neutrino flux is the highest for the magnetospheric model, followed by the current sheet model and the shock model at the last. However, even for the magnetospheric model, the neutrino flux from FRB 200428 is still 2 orders of magnitude below the IceCube sensitivity. The entire FRB population in the universe also only contributes to a small fraction of the neutrino background. Searches for FRB-neutrino associations led to null results and the posted upper limits are still much higher than the model predictions \cite{IcecubeFRB18,luozhang23}.

Stronger neutrino emission may be produced during the birth of magnetars. Ref. \cite{murase09} showed that proton acceleration due to wave dissipation in the magnetar wind when it interacts with the surrounding stellar ejecta. The bright neutrino emission can persist for a few days. Neutrino detection is possible with the current and future neutrino detectors for very nearby new-born magnetars. For extremely nearby sources (e.g. Galactic or local group sources), the MeV thermal neutrinos associated with the birth of a magnetar may be also detected.

\section{Gravitational waves from FRBs and FRB-like events}\label{sec:GWs}

\begin{figure}[th]
\includegraphics[width=5in]{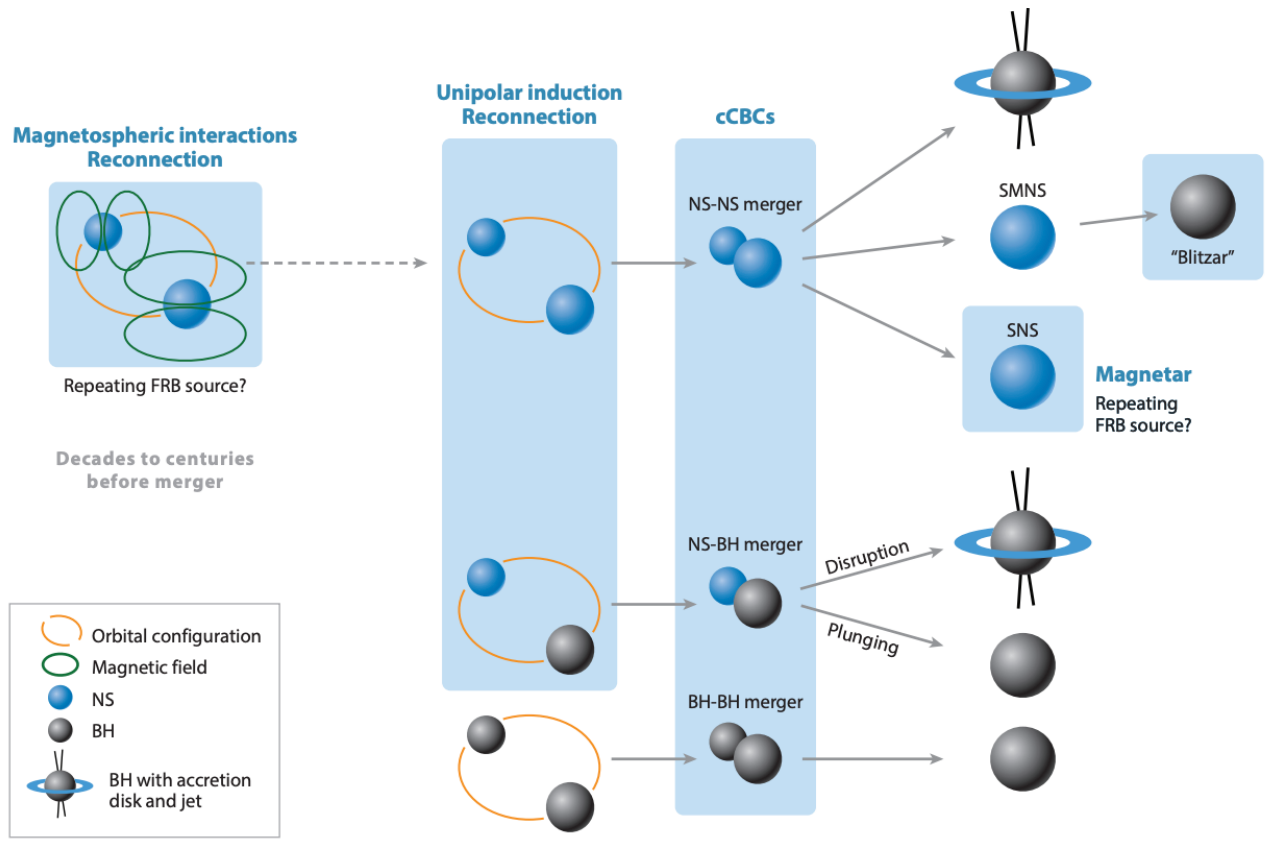}
\caption{Various scenarios to associate FRBs or FRB-like events with gravitational wave sources. Light blue boxes correspond to possible processes to produce FRBs or FRB-like events. Abbreviations: BH, black hole; cCBC: charged compact binary coalescence; FRB: fast radio burst; NS: neutron star; SMNS: supramassive neutron star; SNS: stable neutron star. }
\label{fig:GW-FRBs}
\end{figure}

To be an efficient gravitational wave emitter, the source needs to have a time-varying quadrupole moment. The FRBs as we know them (repeaters from magnetars) are not very good gravitational wave emitters. Nonetheless, many authors have speculated that FRB-like events may be produced around the time of known gravitational wave sources, i.e. CBCs including NS-NS, NS-BH, and even BH-BH mergers. We discuss these topics in this section. The speculations of FRB-like events associated with various CBC events are summarized in Figure \ref{fig:GW-FRBs}.

\subsection{Possible gravitational waves from repeating FRBs}

Within the magnetar framework, emission of FRBs may be associated with crust cracking and weak neutron star tumbling. Such events in principle can generate gravitational waves. The degree of tumbling may not be signficiant. It was suggested \cite{ioka01} that during a magnetar giant flare the magnetar shape may be deformed, which causes a change in the moment of inertia and the total gravitational energy of the system. A gravitational wave strain of $h \sim 5 \times 10^{-21}$ was predicted for a source at 5 kpc, if the gravitational wave energy released is of the order of $10^{47} \ {\rm erg}$. There has been no giant flare in the gravitational wave era to test this optimistic prediction. Upper limits on gravitational emission from some magnetar bursts have been derived using the advanced LIGO and advanced Virgo data \cite{LIGOmagnetar}, which set an upper limit of a few $10^{43} \ {\rm erg}$ gravitational wave energy from these bursts. The X-ray burst associated with the Galactic FRB 20200428D was much less energetic than giant flares \cite{HXMT-SGR,Integral-SGR,AGILE-SGR,Konus-SGR}, and there are observational evidence \cite{tendulkar16} and theoretical arguments \cite{ioka20b} that disfavor the association between FRBs and giant flares. For these reasons, we conclude that there is little chance to detect gravitational waves from any repeating FRBs with the current gravitational wave detectors (LIGO/Virgo/KAGRA, or LVK).

\subsection{FRB-like events from gravitational wave sources}

CBCs are characterized by a gravitational wave chirp signal with the luminosity and frequency rapidly increasing at the end of the merger. If there are any electric or magnetic fields in the merging objects, it is naturally expected that a brief electromagnetic chirp would also accompany  the gravitational wave chirp. Neutron stars carrying magnetospheres would induce additional interactions before the merger, triggering magnetic reconnection that may power brief electromagnetic signals. Whether these signals appear in the form of FRBs or some other fast transients is not clear. In the following, we characterize all these events as FRB-like events and discuss various predictions/speculations associated with the three types of CBC sources the LVK gravitational wave detectors are detecting.

\subsubsection{NS-NS mergers}\label{sec:NS-NS}

NS-NS mergers are known to be associated with short GRBs and kilonovae, as showcased by the example of GW170817/GRB 170817A/AT2017gfo association \cite{GW170817/GRB170817A}. During or shortly after the merger, the system is quite messy, characterized by a dynamically launched neutron-rich ejecta, a central engine (a massive neutron star or a black hole with an accretion disk), and a relativistic jet preferably launched along the axis perpendicular to the orbital plane. Since FRB emission is highly coherent, the emission mechanism is likely delicate and may not survive the dirty environment. Even if an FRB-like event is emitted, it is very difficult to go through the dirty environment without being absorbed. Thus, the FRB-like signals associated with NS-NS mergers should be produced before or much after the merger event itself. Nonetheless, all the processes that generate electromagnetic signals tend to have the highest power at the merger, so many of these scenarios invoke physical mechanisms ``right before'' the merger. We define emission at this phase as the prompt signal, and also discuss other signals much before or after the prompt phase.

\begin{itemize}
\item {\bf Prompt-phase signals:} The most commonly discussed mechanism to generate an FRB-like event right before the merger is through the unipolar inductor effect when a neutron star with a weaker magnetic field penetrating through the magnetic field lines of another neutron star with a stronger field. Such an effect was initially discussed within the context of Jupitor-Io system \cite{goldreich69b}, but has been discussed within the context of NS-NS mergers by various authors \cite{hansen01,piro12,lai12,wang16}. The energy dissipation rate, which could be the electromagnetic emission power, of the system increases rapidly with the shrinking separation $a$ between the two objects with the scaling of $\dot E \propto a^{-7}$. The magnitude may be estimated as \cite{lai12}
\begin{equation}
 L^{\rm uni}_{\rm EM} \sim (10^{44} \ {\rm erg \ s^{-1}}) \left(\frac{B_*}{10^{13} \ {\rm G}} \right)^2 \left(\frac{a}{30 \ {\rm km}} \right)^{-7} \ {\rm erg \ s^{-1}},
\end{equation}
which greatly exceeds the FRB luminosity at the merger. If this process can generate coherent radio emission, then an FRB-like event will be generated even if the FRB has a relatively low radio emission efficiency. The proposed FRB emission mechanism includes magnetospheric curvature radiation mechanism \cite{wang16}, synchrotron maser mechanism invoking relativistic shocks \cite{sridhar21b}, or magnetic reconnection \cite{most23a}. 

A second possible way to generate FRB-like events is through synchronization of the magnetosphere of the post-merger remnant \cite{totani13}. This happens in the merger phase when the dynamical ejecta already blocks a substantial solid angle. To detect an FRB-like event, one either needs a time window or a small solid angle gap from which no ejecta contamination is available and FRB-like events can escape \cite{yamasaki18}.

Finally, there is a generic scenario involving charged CBCs \cite{zhang16a,zhang19} to generate an FRB-like event. This requires that at least one member of the CBC system is globally charged. It is well known that spinning magnets, such as neutron stars, are globally charged \cite{michel82}, so NS-NS and NS-BH mergers are all guaranteed cCBCs. The charged members are continuously accelerated in the orbit, and hence, radiate electric dipole and magnetic dipole radiations with increasing powers towards the merger. To make it more quantitative, it is worth reminding that the gravitational wave luminosity increases as $a^{-5}$, and for an equal-mass pair, the gravitational wave luminosity goes as 
\begin{equation}
 L_{\rm GW} \simeq \frac{1}{5} \frac{c^5}{G} \left(\frac{r_s}{a}\right)^5,
\end{equation}
where $c^5/G \simeq 3.6\times 10^{59} \ {\rm erg \ s^{-1}}$ is a maximum luminosity defined by fundamental constants, and $r_s = 2 G m /c^2$ is the Schwarzschild radius of each merging member with mass $m$. Now consider that each mass carries a dimensionless charge of $\hat q = Q/Q_c$, where $Q_c \equiv 2\sqrt{G} m$ is the characteristic charge a merging member defined by its mass $m$ (at $Q_c$ the spacetime would be significantly curved), again for identical merger members with the same $m$ and $\hat q$, the electromagnetic luminosity due to electric dipole radiation and magnetic dipole radiation can be characterized by \cite{zhang19}
\begin{equation}
 L_{\rm EM}^{\rm E-dip} \simeq \frac{1}{6} \frac{c^5}{G} \hat q^2 \left(\frac{r_s}{a}\right)^4, ~~~~
 L_{\rm EM}^{\rm B-dip} \simeq \frac{196}{1875}
 \frac{c^5}{G} \hat q^2 \left(\frac{r_s}{a}\right)^{15},
\end{equation}
which scales as $\propto a^{-4}$ and $\propto a^{-15}$, respectively. For a characteristic charge number $\hat q \sim 10^{-7}$ relevant for the Crab pulsar, the electric dipole radiation power can reach $\sim 5\times 10^{42} \ {\rm erg \ s^{-1}}$. Taking a reasonably small efficiency to the radio band, such FRB-like events are quite weak, with a power comparable to that of the Galactic FRB 20200428D. 
\item {\bf Before the merger:} The magnetospheres of two neutron stars start to interact long before the merger. Even though most authors believe that significant energy dissipation happens shortly before the coalescence \cite{hansen01,gourgouliatos19}, Ref. \cite{zhang20} argued that FRB-like events may be generated decades even centuries before the merger due to the sporadic magnetic reconnection triggered by the interacting magnetospheres. The isotropic equivalent luminosity of these events can reach that of repeating FRBs if the emitters are narrowly beamed, which is likely the case for the jets ejected from magnetic reconnection in a highly magnetized (high-$\sigma$) plasma. According to this scenario, some active repeaters could be interacting neutron stars that are close to merging, and some nearby systems may be detected as gravitational wave sources by future space-borne gravitational wave detectors such as LISA, Taiji and TianQin.
\item {\bf After the merger:} If the merging neutron stars are light enough and the neutron star equation of state is stiff enough, an NS-NS merger may leave behind a long-lived supramassive neutron star that can survive for an extended period of time through rigid rotation before collapsing to the black hole (i.e. an SMNS). Under extreme conditions (very light neutron star members in the system), a merger may even lead to a massive stable neutron star (SNS) that will never collapse. In both cases, late-time FRB or FRB-like emissions are likely. 

For the former scenario, an FRB-like event may be emitted when the SMNS collapses to a black hole with the magnetosphere ejected in a ``blitzar'' event \cite{falcke14,most18}. For NS-NS mergers, such a collapse may happen several hundreds of seconds after the merger, as indicated by the duration of the X-ray ``internal plateaus'' observed in nearly half of short GRBs \cite{rowlinson13,lv15}. For this reason, Ref. \cite{zhang14} suggested that an FRB may follow a short GRB with a delay timescale of minutes to hours. The dirty NS-NS merger environment is a concern regarding FRB propagation. However, along the direction of the short GRB jet which is a clean funnel cleared by the relativistic ejecta, an FRB-like event can propagate unimpeded \cite{deng14}. So in the scenario of Ref. \cite{zhang14}, one expects a GW-sGRB-FRB tripple association. 

For the latter scenario, a long-lived SNS may be able to emit repeating FRBs using the standard magnetar engine if the post-merger remnant reaches the magnetar magnetic field strength \cite{margalit19,wang20}. In this case, a repeating FRB source is preceded by an NS-NS merger gravitational wave event.
\end{itemize}

\subsubsection{NS-BH mergers}

NS-BH mergers have two types. If the mass ratio between the BH and the NS, $q=m_{\rm BH}/m_{\rm NS} > 1$, is not too large and/or the BH has a high prograde spin, the neutron star is likely tidally disrupted before the merger. The merged black hole is likely surrounded by neutron-rich matter in the forms of dynamical ejecta, accretion disk, and possible a relativistic jet. The situation is very similar to the NS-NS merger case. 
Since the post-merger product is a black hole, no magnetar-related post-merger FRB emission is possible. The FRB-like events in this scenario can be promptly produced through the unipolar induction or the cCBC mechanisms. The radio waves generated are subject to the absorption effect from the messy environment. 

The second type of NS-BH merger is called a ``plunging event''. This applies to the case of $q \gg 1$ and/or a large retrograde spin of the BH. The NS tidal disruption radius is smaller than the BH horizon so that the entire NS is swallowed by the BH without any material left outside the horizon. Such a case has a clean environment and is very suitable for the propagation of FRB-like events (if they are produced). A guaranteed signal is the cCBC signal discussed in \S\ref{sec:NS-NS}. Alternatively, the magnetic field of the NS may thread through the BH before the merger, leading to magnetic field twisting and eventually reconnection. This would also lead to one or more reconnection-triggered sub-FRB-like-events on top of the electromagnetic chirp signal discussed above \cite{most23b}.

Another related mechanism is that a BH can become charged when traveling in the magnetic field of the neutron star \cite{wald74}. Such a black hole battery mechanism has been suggested to generate FRB-like events for plunging NS-BH merger systems \cite{mingarelli15,levin18,dai19}. 

\subsubsection{BH-BH mergers}

Finally, BH-BH mergers are usually electromagnetically quiet. A weak cCBC signal is possible if one of the BH is charged \cite{zhang16a,liebling16,liu16,deng18}. A key question is whether BHs could be charged and how long the charge could sustain. So far, no robust detection of electromagnetic signals from BH-BH mergers has been made. The non-detection could pose some loose upper limits on the BH charge \cite{klingler21}.

\subsection{Case studies}

Searches of gravitational wave - FRB associations have been carried out and so far only upper limits are obtained during the prompt phase \cite{LIGO-FRB22}. Nonetheless, a putative association has been reported in the literature \cite{moroianu22,panther23}. A related case of a short GRB - FRB association was also reported \cite{rowlinson23}. We discuss them in detail below. 

\subsubsection{The putative GW190425/FRB 20190425A association}

FRB 20190425A was detected by the CHIME telescope and occurred about 2.5 hours after GW190425, a likely BNS merger event. Its location is within the GW190425's sky localization area. Its DM-inferred distance is consistent with that of GW190425 with a small chance probability. Overall, the significance of the association is below $3\sigma$. The FRB is a one-off, single-peak, bright burst within the CHIME's FRB catalog. It does not possess the properties of repeating FRBs (e.g. broad width, narrow spectrum, spectral down-drifting) and looks like a genuinely non-repeating FRB. This association may be interpreted within the framework of SMNS collapse after 2.5 hours as outlined in \cite{zhang14}. Within this scenario, a short GRB is needed to clear the funnel to allow the FRB to propagate. There was a $\gamma$-ray excess detected by INTEGRAL that might be consistent with a sGRB, even though the event could not be confirmed. If the association is astronomical, then the required neutron star equation of state is very stiff, since the GW190425 BNS system already contains a massive neutron star.

A few host galaxy candidates were identified by \cite{panther23}, with the most probable candidate being UGC10667 at redshift $z=0.03136 \pm 0.00009$. However, this host galaxy raises an inconsistency to the scenario. As pointed out by \cite{bhardwaj23}, at the distance of that redshift, the gravitational wave would be too strong compared with the detected level if the system is face-on. In order to satisfy the GW observational constraint, the system needs to have a viewing angle greater than 30 degree with a $>99.99\%$ confidence level. This makes the association unlikely because the viewing direction would be opaque for the FRB. Another issue for the association is that there is a tight kilonova upper limit which is inconsistent with the expectation of a magnetar-powered bright kilonova \cite{smartt23,radice23}. All these cast doubt on this specific association. 

\subsubsection{The putative GRB 201006A/LOFAR radio transient association}

\cite{rowlinson23} reported the detection of a $5.6\sigma$, short, coherent radio flash at 144 MHz with the LOFAR telescope at 76.6 min after the short-duration GRB 201006A. The GRB was a {\em Swift}-detected short hard GRB without redshift measurement. Dispersion analysis of the radio burst gave DM $\sim 770 \ {\rm pc \ cm^{-3}}$, corresponding to $z \sim 0.58$ based on the ${\rm DM_{IGM}} - z$ relation, which is consistent with the $z$-distribution of short GRBs. According to the standard model of short GRBs, GRB 201006A could have been associated with gravitational waves due to NS-NS merger, but this redshift is far outside of the horizon of LVK GW detectors for BNS mergers. There was a small (27 arcsecond) offset between the radio flash position and the GRB location, but the authors suggested a physical association based on a small chance probability argument. Such an association, if physical, is again consistent with the post-merger FRB model as suggested in \cite{zhang14}. However, because of the low signficance of the association, more similar cases are needed to robustly establish an association between BNS mergers/GWs and a special type of FRBs. 

\section{Summary and prospects}\label{sec:summary}

The origin of cosmological FRBs is still mysterious. Multi-wavelength, multi-messenger signals associated with any of these sources would shed light on their possible origins. However, other than the Galactic FRB 20200428D that was associated with a moderately bright magnetar X-ray burst, so far no robust associations in other wavelengths or other multi-messenger channels have been discovered, both during the prompt phase and before or after the bursts. Such non-detection is expected, as FRBs are intrinsically faint cosmological objects and the predicted levels of multi-wavelength, multi-messenger emissions are (much) lower than the sensitivities of the current telescopes or detectors. It is foreseen that major progress can be only made for very rare nearby events (like the Galactic FRB 20200428D), or if some of the FRBs are indeed associated with catastrophic events such as CBC gravitational wave events. The non-detections, especially with ever more stringent upper limits, would help to eliminate some models and improve our understanding of the physics of FRBs.

\bigskip
Acknowledgement: The author thanks all the colleagues who have had fruitful collaborations or insightful discussions with him to understand the origin of FRBs. Special thanks go to Mohit Bhardwaj and Sibasish Laha who read the proof version and helped to finalize the paper..  

%\acknowledgement The author acknowledges many colleagues for collaborations and discussions

%\section*{REFERENCES CITED}

%\bibliographystyle{natbib}
%\bibliography{FRB}

%\bibliographystyle{naturemag}
%\bibliographystyle{ar-style5}
%\bibliography{FRB}

\begin{thebibliography}{202}
\expandafter\ifx\csname natexlab\endcsname\relax\def\natexlab#1{#1}\fi

\bibitem{lorimer07}
{Lorimer} DR, et~al.
\newblock \textit{Science} 318:777 (2007)

\bibitem{thornton13}
{Thornton} D, et~al.
\newblock \textit{Science} 341:53 (2013)

\bibitem{CHIME-SGR}
{CHIME/FRB Collaboration}, et~al.
\newblock \textit{\nat} 587:54 (2020)

\bibitem{STARE2-SGR}
{Bochenek} CD, et~al.
\newblock \textit{\nat} 587:59 (2020)

\bibitem{HXMT-SGR}
{Li} CK, et~al.
\newblock \textit{Nature Astronomy} 5:378 (2021)

\bibitem{Integral-SGR}
{Mereghetti} S, et~al.
\newblock \textit{\apjl} 898:L29 (2020)

\bibitem{Konus-SGR}
{Ridnaia} A, et~al.
\newblock \textit{Nature Astronomy} 5:372 (2021)

\bibitem{AGILE-SGR}
{Tavani} M, et~al.
\newblock \textit{Nature Astronomy} 5:401 (2021)

\bibitem{katz18}
{Katz} JI.
\newblock \textit{Progress in Particle and Nuclear Physics} 103:1 (2018)

\bibitem{popov18}
{Popov} SB, {Postnov} KA, {Pshirkov} MS.
\newblock \textit{Physics Uspekhi} 61:965 (2018)

\bibitem{platts19}
{Platts} E, et~al.
\newblock \textit{\physrep} 821:1 (2019)

\bibitem{petroff19}
{Petroff} E, {Hessels} JWT, {Lorimer} DR.
\newblock \textit{\aapr} 27:4 (2019)

\bibitem{petroff22}
{Petroff} E, {Hessels} JWT, {Lorimer} DR.
\newblock \textit{\aapr} 30:2 (2022)

\bibitem{cordes19}
{Cordes} JM, {Chatterjee} S.
\newblock \textit{\araa} 57:417 (2019)

\bibitem{zhang20b}
{Zhang} B.
\newblock \textit{\nat} 587:45 (2020)

\bibitem{xiao21}
{Xiao} D, {Wang} F, {Dai} Z.
\newblock \textit{Science China Physics, Mechanics, and Astronomy} 64:249501
  (2021)

\bibitem{lyubarsky21}
{Lyubarsky} Y.
\newblock \textit{Universe} 7:56 (2021)

\bibitem{bailes22}
{Bailes} M.
\newblock \textit{Science} 378:abj3043 (2022)

\bibitem{zhang23}
{Zhang} B.
\newblock \textit{Reviews of Modern Physics} 95:035005 (2023)

\bibitem{chime-1st-catalog}
{The CHIME/FRB Collaboration}, et~al.
\newblock \textit{arXiv e-prints} :arXiv:2106.04352 (2021)

\bibitem{zhou22}
{Zhou} DJ, et~al.
\newblock \textit{Research in Astronomy and Astrophysics} 22:124001 (2022)

\bibitem{champion16}
{Champion} DJ, et~al.
\newblock \textit{\mnras} 460:L30 (2016)

\bibitem{pleunis21b}
{Pleunis} Z, et~al.
\newblock \textit{\apj} 923:1 (2021)

\bibitem{hessels19}
{Hessels} JWT, et~al.
\newblock \textit{\apjl} 876:L23 (2019)

\bibitem{chime-repeaters}
{CHIME/FRB Collaboration}, et~al.
\newblock \textit{\apjl} 885:L24 (2019)

\bibitem{chime-new-repeaters}
{Chime/Frb Collaboration}, et~al.
\newblock \textit{\apj} 947:83 (2023)

\bibitem{zhangy18}
{Zhang} YG, et~al.
\newblock \textit{\apj} 866:149 (2018)

\bibitem{xuh22}
{Xu} H, et~al.
\newblock \textit{\nat} 609:685 (2022)

\bibitem{niujr22}
{Niu} JR, et~al.
\newblock \textit{Research in Astronomy and Astrophysics} 22:124004 (2022)

\bibitem{chime-period}
{Chime/Frb Collaboration} Andersen BC, et~al.
\newblock \textit{\nat} 607:256 (2022)

\bibitem{chime-periodic}
{Chime/Frb Collaboration}, et~al.
\newblock \textit{\nat} 582:351 (2020)

\bibitem{rajwade20}
{Rajwade} KM, et~al.
\newblock \textit{\mnras}  (2020)

\bibitem{zhangyk23}
{Zhang} YK, et~al.
\newblock \textit{\apj} 955:142 (2023)

\bibitem{michilli18}
{Michilli} D, et~al.
\newblock \textit{\nat} 553:182 (2018)

\bibitem{jiangjc22}
{Jiang} JC, et~al.
\newblock \textit{Research in Astronomy and Astrophysics} 22:124003 (2022)

\bibitem{kumar22}
{Kumar} P, et~al.
\newblock \textit{\mnras} 512:3400 (2022)

\bibitem{day20}
{Day} CK, et~al.
\newblock \textit{\mnras} 497:3335 (2020)

\bibitem{luo20b}
{Luo} R, et~al.
\newblock \textit{\nat} 586:693 (2020)

\bibitem{macquart20}
{Macquart} JP, et~al.
\newblock \textit{\nat} 581:391 (2020)

\bibitem{tendulkar17}
{Tendulkar} SP, et~al.
\newblock \textit{\apjl} 834:L7 (2017)

\bibitem{bannister19}
{Bannister} KW, et~al.
\newblock \textit{Science} 365:565 (2019)

\bibitem{ravi19}
{Ravi} V, et~al.
\newblock \textit{\nat} 572:352 (2019)

\bibitem{bhandari20}
{Bhandari} S, et~al.
\newblock \textit{\apjl} 895:L37 (2020)

\bibitem{lizhang20}
{Li} Y, {Zhang} B.
\newblock \textit{\apjl} 899:L6 (2020)

\bibitem{dong23}
{Dong} Y, et~al.
\newblock \textit{arXiv e-prints} :arXiv:2307.06995 (2023)

\bibitem{bhardwaj21}
{Bhardwaj} M, et~al.
\newblock \textit{\apjl} 910:L18 (2021)

\bibitem{kirsten22}
{Kirsten} F, et~al.
\newblock \textit{\nat} 602:585 (2022)

\bibitem{anna-thomas23}
{Anna-Thomas} R, et~al.
\newblock \textit{Science} 380:599 (2023)

\bibitem{lu20}
{Lu} W, {Kumar} P, {Zhang} B.
\newblock \textit{\mnras} 498:1397 (2020)

\bibitem{ravi19b}
{Ravi} V.
\newblock \textit{Nature Astronomy} 3:928 (2019)

\bibitem{luo20}
{Luo} R, et~al.
\newblock \textit{\mnras} 494:665 (2020)

\bibitem{luo18}
{Luo} R, {Lee} K, {Lorimer} DR, {Zhang} B.
\newblock \textit{\mnras} 481:2320 (2018)

\bibitem{lu18}
{Lu} W, {Kumar} P.
\newblock \textit{\mnras} 477:2470 (2018)

\bibitem{shin23}
{Shin} K, et~al.
\newblock \textit{\apj} 944:105 (2023)

\bibitem{zhangzhang22}
{Zhang} RC, {Zhang} B.
\newblock \textit{\apjl} 924:L14 (2022)

\bibitem{hashimoto22}
{Hashimoto} T, et~al.
\newblock \textit{\mnras} 511:1961 (2022)

\bibitem{qiang22}
{Qiang} DC, {Li} SL, {Wei} H.
\newblock \textit{\jcap} 2022:040 (2022)

\bibitem{popov10}
{Popov} SB, {Postnov} KA. 2010.
\newblock In \textit{Evolution of Cosmic Objects through their Physical
  Activity}, eds. HA~{Harutyunian}, AM~{Mickaelian}, Y~{Terzian}

\bibitem{kulkarni14}
{Kulkarni} SR, et~al.
\newblock \textit{\apj} 797:70 (2014)

\bibitem{katz16}
{Katz} JI.
\newblock \textit{\apj} 826:226 (2016)

\bibitem{lyubarsky14}
{Lyubarsky} Y.
\newblock \textit{\mnras} 442:L9 (2014)

\bibitem{metzger17}
{Metzger} BD, {Berger} E, {Margalit} B.
\newblock \textit{\apj} 841:14 (2017)

\bibitem{kumar17}
{Kumar} P, {Lu} W, {Bhattacharya} M.
\newblock \textit{\mnras} 468:2726 (2017)

\bibitem{beloborodov17}
{Beloborodov} AM.
\newblock \textit{\apjl} 843:L26 (2017)

\bibitem{yangzhang18}
{Yang} YP, {Zhang} B.
\newblock \textit{\apj} 868:31 (2018)

\bibitem{metzger19}
{Metzger} BD, {Margalit} B, {Sironi} L.
\newblock \textit{\mnras} 485:4091 (2019)

\bibitem{wadiasingh19}
{Wadiasingh} Z, {Timokhin} A.
\newblock \textit{\apj} 879:4 (2019)

\bibitem{beloborodov20}
{Beloborodov} AM.
\newblock \textit{\apj} 896:142 (2020)

\bibitem{margalit20}
{Margalit} B, {Beniamini} P, {Sridhar} N, {Metzger} BD.
\newblock \textit{\apjl} 899:L27 (2020)

\bibitem{lyutikov21}
{Lyutikov} M.
\newblock \textit{\apj} 922:166 (2021)

\bibitem{yangzhang21}
{Yang} YP, {Zhang} B.
\newblock \textit{\apj} 919:89 (2021)

\bibitem{zhang22}
{Zhang} B.
\newblock \textit{\apj} 925:53 (2022)

\bibitem{cordes16}
{Cordes} JM, {Wasserman} I.
\newblock \textit{\mnras} 457:232 (2016)

\bibitem{connor16}
{Connor} L, {Sievers} J, {Pen} UL.
\newblock \textit{\mnras} 458:L19 (2016)

\bibitem{zhang17}
{Zhang} B.
\newblock \textit{\apjl} 836:L32 (2017)

\bibitem{dai16}
{Dai} ZG, {Wang} JS, {Wu} XF, {Huang} YF.
\newblock \textit{\apj} 829:27 (2016)

\bibitem{katz17b}
{Katz} JI.
\newblock \textit{\mnras} 471:L92 (2017)

\bibitem{sridhar21}
{Sridhar} N, et~al.
\newblock \textit{\apj} 917:13 (2021)

\bibitem{totani13}
{Totani} T.
\newblock \textit{\pasj} 65:L12 (2013)

\bibitem{mingarelli15}
{Mingarelli} CMF, {Levin} J, {Lazio} TJW.
\newblock \textit{\apjl} 814:L20 (2015)

\bibitem{zhang16a}
{Zhang} B.
\newblock \textit{\apjl} 827:L31 (2016)

\bibitem{wang16}
{Wang} JS, et~al.
\newblock \textit{\apjl} 822:L7 (2016)

\bibitem{levin18}
{Levin} J, {D'Orazio} DJ, {Garcia-Saenz} S.
\newblock \textit{\prd} 98:123002 (2018)

\bibitem{zhang19}
{Zhang} B.
\newblock \textit{\apjl} 873:L9 (2019)

\bibitem{dai19}
{Dai} ZG.
\newblock \textit{\apjl} 873:L13 (2019)

\bibitem{falcke14}
{Falcke} H, {Rezzolla} L.
\newblock \textit{\aap} 562:A137 (2014)

\bibitem{zhang14}
{Zhang} B.
\newblock \textit{\apjl} 780:L21 (2014)

\bibitem{most18}
{Most} ER, {Nathanail} A, {Rezzolla} L.
\newblock \textit{\apj} 864:117 (2018)

\bibitem{kremer21}
{Kremer} K, {Piro} AL, {Li} D.
\newblock \textit{\apjl} 917:L11 (2021)

\bibitem{lu22}
{Lu} W, {Beniamini} P, {Kumar} P.
\newblock \textit{\mnras} 510:1867 (2022)

\bibitem{kremer23}
{Kremer} K, et~al.
\newblock \textit{\apj} 944:6 (2023)

\bibitem{kumar20a}
{Kumar} P, {Bo{\v{s}}njak} {\v{Z}}.
\newblock \textit{\mnras} 494:2385 (2020)

\bibitem{lyubarsky20}
{Lyubarsky} Y.
\newblock \textit{\apj} 897:1 (2020)

\bibitem{qu22b}
{Qu} Y, {Kumar} P, {Zhang} B.
\newblock \textit{\mnras} 515:2020 (2022)

\bibitem{qu23}
{Qu} Y, {Zhang} B, {Kumar} P.
\newblock \textit{\mnras} 518:66 (2023)

\bibitem{sironi21}
{Sironi} L, {Plotnikov} I, {N{\"a}ttil{\"a}} J, {Beloborodov} AM.
\newblock \textit{\prl} 127:035101 (2021)

\bibitem{wangwy22}
{Wang} WY, et~al.
\newblock \textit{\apj} 927:105 (2022)

\bibitem{wangwy22b}
{Wang} WY, et~al.
\newblock \textit{\mnras} 517:5080 (2022)

\bibitem{quzhang23}
{Qu} Y, {Zhang} B.
\newblock \textit{\mnras} 522:2448 (2023)

\bibitem{lid21}
{Li} D, et~al.
\newblock \textit{\nat} 598:267 (2021)

\bibitem{zhangyk22}
{Zhang} YK, et~al.
\newblock \textit{Research in Astronomy and Astrophysics} 22:124002 (2022)

\bibitem{beloborodov21}
{Beloborodov} AM.
\newblock \textit{\apjl} 922:L7 (2021)

\bibitem{beloborodov23}
{Beloborodov} AM.
\newblock \textit{arXiv e-prints} :arXiv:2307.12182 (2023)

\bibitem{lyutikov23}
{Lyutikov} M.
\newblock \textit{arXiv e-prints} :arXiv:2310.01177 (2023)

\bibitem{lin20}
{Lin} L, et~al.
\newblock \textit{\nat} 587:63 (2020)

\bibitem{margalit19}
{Margalit} B, {Berger} E, {Metzger} BD.
\newblock \textit{\apj} 886:110 (2019)

\bibitem{wang20}
{Wang} FY, et~al.
\newblock \textit{\apj} 891:72 (2020)

\bibitem{zhang20}
{Zhang} B.
\newblock \textit{\apjl} 890:L24 (2020)

\bibitem{piro12}
{Piro} AL.
\newblock \textit{\apj} 755:80 (2012)

\bibitem{lai12}
{Lai} D.
\newblock \textit{\apjl} 757:L3 (2012)

\bibitem{liebling16}
{Liebling} SL, {Palenzuela} C.
\newblock \textit{\prd} 94:064046 (2016)

\bibitem{deng18}
{Deng} CM, {Cai} Y, {Wu} XF, {Liang} EW.
\newblock \textit{\prd} 98:123016 (2018)

\bibitem{liu16}
{Liu} T, {Romero} GE, {Liu} ML, {Li} A.
\newblock \textit{\apj} 826:82 (2016)

\bibitem{fraschetti18}
{Fraschetti} F.
\newblock \textit{\jcap} 2018:054 (2018)

\bibitem{luojw23}
{Luo} JW, {Zhu-Ge} JM, {Zhang} B.
\newblock \textit{\mnras} 518:1629 (2023)

\bibitem{tendulkar16}
{Tendulkar} SP, {Kaspi} VM, {Patel} C.
\newblock \textit{\apj} 827:59 (2016)

\bibitem{yangyh21}
{Yang} YH, et~al.
\newblock \textit{\apjl} 906:L12 (2021)

\bibitem{younes21}
{Younes} G, et~al.
\newblock \textit{Nature Astronomy} 5:408 (2021)

\bibitem{lixb22}
{Li} X, et~al.
\newblock \textit{\apj} 931:56 (2022)

\bibitem{zhang21}
{Zhang} B.
\newblock \textit{\apjl} 907:L17 (2021)

\bibitem{chenzhang22}
{Chen} CJ, {Zhang} B.
\newblock \textit{\mnras} 519:6284 (2023)

\bibitem{ioka20b}
{Ioka} K.
\newblock \textit{\apjl} 904:L15 (2020)

\bibitem{ge23}
{Ge} MY, et~al.
\newblock \textit{\apj} 953:67 (2023)

\bibitem{scholz17}
{Scholz} P, et~al.
\newblock \textit{\apj} 846:80 (2017)

\bibitem{scholz20}
{Scholz} P, et~al.
\newblock \textit{\apj} 901:165 (2020)

\bibitem{laha22a}
{Laha} S, et~al.
\newblock \textit{\apj} 930:172 (2022)

\bibitem{laha22b}
{Laha} S, et~al.
\newblock \textit{\apj} 929:173 (2022)

\bibitem{piro21}
{Piro} L, et~al.
\newblock \textit{\aap} 656:L15 (2021)

\bibitem{pearlman23}
{Pearlman} AB, et~al.
\newblock \textit{arXiv e-prints} :arXiv:2308.10930 (2023)

\bibitem{bannister12}
{Bannister} KW, {Murphy} T, {Gaensler} BM, {Reynolds} JE.
\newblock \textit{\apj} 757:38 (2012)

\bibitem{delaunay16}
{DeLaunay} JJ, et~al.
\newblock \textit{\apjl} 832:L1 (2016)

\bibitem{cunningham19}
{Cunningham} V, et~al.
\newblock \textit{\apj} 879:40 (2019)

\bibitem{martone19}
{Martone} R, et~al.
\newblock \textit{\aap} 631:A62 (2019)

\bibitem{anumarlapudi20}
{Anumarlapudi} A, {Bhalerao} V, {Tendulkar} SP, {Balasubramanian} A.
\newblock \textit{\apj} 888:40 (2020)

\bibitem{casentini20}
{Casentini} C, et~al.
\newblock \textit{\apjl} 890:L32 (2020)

\bibitem{guidorzi20}
{Guidorzi} C, et~al.
\newblock \textit{\aap} 642:A160 (2020)

\bibitem{tohuvavohu20}
{Tohuvavohu} A, et~al.
\newblock \textit{\apj} 900:35 (2020)

\bibitem{mereghetti21}
{Mereghetti} S, {Topinka} M, {Rigoselli} M, {G{\"o}tz} D.
\newblock \textit{\apjl} 921:L3 (2021)

\bibitem{verrecchia21}
{Verrecchia} F, et~al.
\newblock \textit{\apj} 915:102 (2021)

\bibitem{curtin23}
{Curtin} AP, et~al.
\newblock \textit{\apj} 954:154 (2023)

\bibitem{zhangzhang17}
{Zhang} BB, {Zhang} B.
\newblock \textit{\apjl} 843:L13 (2017)

\bibitem{yangyh19}
{Yang} YH, {Zhang} BB, {Zhang} B.
\newblock \textit{\apjl} 875:L19 (2019)

\bibitem{yang19}
{Yang} YP, {Zhang} B, {Wei} JY.
\newblock \textit{\apj} 878:89 (2019)

\bibitem{yangyp21}
{Yang} YP.
\newblock \textit{\apj} 920:34 (2021)

\bibitem{ioka20}
{Ioka} K, {Zhang} B.
\newblock \textit{\apjl} 893:L26 (2020)

\bibitem{lyutikov20}
{Lyutikov} M, {Barkov} MV, {Giannios} D.
\newblock \textit{\apjl} 893:L39 (2020)

\bibitem{wangfy22}
{Wang} FY, {Zhang} GQ, {Dai} ZG, {Cheng} KS.
\newblock \textit{Nature Communications} 13:4382 (2022)

\bibitem{zhu23}
{Zhu} W, et~al.
\newblock \textit{Science Advances} 9:eadf6198 (2023)

\bibitem{hardy17}
{Hardy} LK, et~al.
\newblock \textit{\mnras} 472:2800 (2017)

\bibitem{MAGIC18}
{MAGIC Collaboration}, et~al.
\newblock \textit{\mnras} 481:2479 (2018)

\bibitem{andreoni20}
{Andreoni} I, et~al.
\newblock \textit{\apjl} 896:L2 (2020)

\bibitem{niino22}
{Niino} Y, et~al.
\newblock \textit{\apj} 931:109 (2022)

\bibitem{marcote20}
{Marcote} B, et~al.
\newblock \textit{\nat} 577:190 (2020)

\bibitem{kilpatrick21}
{Kilpatrick} CD, et~al.
\newblock \textit{\apjl} 907:L3 (2021)

\bibitem{hiramatsu23}
{Hiramatsu} D, et~al.
\newblock \textit{\apjl} 947:L28 (2023)

\bibitem{pleunis21}
{Pleunis} Z, et~al.
\newblock \textit{\apjl} 911:L3 (2021)

\bibitem{gajjar18}
{Gajjar} V, et~al.
\newblock \textit{\apj} 863:2 (2018)

\bibitem{meszarosrees97}
{M{\'e}sz{\'a}ros} P, {Rees} MJ.
\newblock \textit{\apj} 476:232 (1997)

\bibitem{sari98}
{Sari} R, {Piran} T, {Narayan} R.
\newblock \textit{\apjl} 497:L17 (1998)

\bibitem{yi14}
{Yi} SX, {Gao} H, {Zhang} B.
\newblock \textit{\apjl} 792:L21 (2014)

\bibitem{taylor05}
{Taylor} GB, et~al.
\newblock \textit{\apjl} 634:L93 (2005)

\bibitem{kaspi17}
{Kaspi} VM, {Beloborodov} AM.
\newblock \textit{\araa} 55:261 (2017)

\bibitem{law19}
{Law} CJ, et~al.
\newblock \textit{\apj} 886:24 (2019)

\bibitem{men19}
{Men} Y, et~al.
\newblock \textit{\mnras} 489:3643 (2019)

\bibitem{madison19}
{Madison} DR, et~al.
\newblock \textit{\apj} 887:252 (2019)

\bibitem{rowlinson19}
{Rowlinson} A, {Anderson} GE.
\newblock \textit{\mnras} 489:3316 (2019)

\bibitem{bouwhuis20}
{Bouwhuis} M, et~al.
\newblock \textit{\mnras} 497:125 (2020)

\bibitem{palliyaguru21}
{Palliyaguru} NT, et~al.
\newblock \textit{\mnras} 501:541 (2021)

\bibitem{wangxg20}
{Wang} XG, et~al.
\newblock \textit{\apjl} 894:L22 (2020)

\bibitem{lil22}
{Li} L, et~al.
\newblock \textit{\apj} 929:139 (2022)

\bibitem{chatterjee17}
{Chatterjee} S, et~al.
\newblock \textit{\nat} 541:58 (2017)

\bibitem{niu22}
{Niu} CH, et~al.
\newblock \textit{\nat} 606:873 (2022)

\bibitem{bruni23}
{Bruni} G et~al.
\newblock \textit{arXiv e-prints} :arXiv:2312.15296 (2023)

\bibitem{yang16}
{Yang} YP, {Zhang} B, {Dai} ZG.
\newblock \textit{\apjl} 819:L12 (2016)

\bibitem{murase16}
{Murase} K, {Kashiyama} K, {M\'esz\'aros} P.
\newblock \textit{\mnras} 461:1498 (2016)

\bibitem{sridhar22}
{Sridhar} N, {Metzger} B.
\newblock \textit{\apj} 937:5 (2022)

\bibitem{yang20b}
{Yang} YP, {LI} QC, {Zhang} B.
\newblock \textit{\apj} 895:7 (2020)

\bibitem{yang22}
{Yang} YP, et~al.
\newblock \textit{\apjl} 928:L16 (2022)

\bibitem{law22}
{Law} CJ, {Connor} L, {Aggarwal} K.
\newblock \textit{\apj} 927:55 (2022)

\bibitem{zhang03}
{Zhang} B, et~al.
\newblock \textit{\apj} 595:346 (2003)

\bibitem{metzger20}
{Metzger} BD, {Fang} K, {Margalit} B.
\newblock \textit{\apjl} 902:L22 (2020)

\bibitem{qu22a}
{Qu} Y, {Zhang} B.
\newblock \textit{\mnras} 511:972 (2022)

\bibitem{IcecubeFRB18}
{Aartsen} MG, et~al.
\newblock \textit{\apj} 857:117 (2018)

\bibitem{luozhang23}
{Luo} JW, {Zhang} B.
\newblock \textit{arXiv e-prints} :arXiv:2112.11375 (2021)

\bibitem{murase09}
{Murase} K, {M{\'e}sz{\'a}ros} P, {Zhang} B.
\newblock \textit{\prd} 79:103001 (2009)

\bibitem{ioka01}
{Ioka} K.
\newblock \textit{\mnras} 327:639 (2001)

\bibitem{LIGOmagnetar}
{The LIGO Scientific Collaboration}, et~al.
\newblock \textit{arXiv e-prints} :arXiv:2210.10931 (2022)

\bibitem{GW170817/GRB170817A}
{Abbott} BP, et~al.
\newblock \textit{\apjl} 848:L12 (2017)

\bibitem{goldreich69b}
{Goldreich} P, {Lynden-Bell} D.
\newblock \textit{\apj} 156:59 (1969)

\bibitem{hansen01}
{Hansen} BMS, {Lyutikov} M.
\newblock \textit{\mnras} 322:695 (2001)

\bibitem{sridhar21b}
{Sridhar} N, et~al.
\newblock \textit{\mnras} 501:3184 (2021)

\bibitem{most23a}
{Most} ER, {Philippov} AA.
\newblock \textit{\prl} 130:245201 (2023)

\bibitem{yamasaki18}
{Yamasaki} S, {Totani} T, {Kiuchi} K.
\newblock \textit{\pasj} 70:39 (2018)

\bibitem{michel82}
{Michel} FC.
\newblock \textit{Reviews of Modern Physics} 54:1 (1982)

\bibitem{gourgouliatos19}
{Gourgouliatos} KN, {Lynden-Bell} D.
\newblock \textit{\mnras} 482:1942 (2019)

\bibitem{rowlinson13}
{Rowlinson} A, et~al.
\newblock \textit{\mnras} 430:1061 (2013)

\bibitem{lv15}
{L{\"u}} HJ, et~al.
\newblock \textit{\apj} 805:89 (2015)

\bibitem{deng14}
{Deng} W, {Zhang} B.
\newblock \textit{\apjl} 783:L35 (2014)

\bibitem{most23b}
{Most} ER, {Philippov} AA.
\newblock \textit{\apjl} 956:L33 (2023)

\bibitem{wald74}
{Wald} RM.
\newblock \textit{\prd} 10:1680 (1974)

\bibitem{klingler21}
{Klingler} NJ, et~al.
\newblock \textit{\apj} 907:97 (2021)

\bibitem{LIGO-FRB22}
{The LIGO Scientific Collaboration}, et~al.
\newblock \textit{arXiv e-prints} :arXiv:2203.12038 (2022)

\bibitem{moroianu22}
{Moroianu} A, et~al.
\newblock \textit{Nature Astronomy} 7:579 (2023)

\bibitem{panther23}
{Panther} FH, et~al.
\newblock \textit{\mnras} 519:2235 (2023)

\bibitem{rowlinson23}
{Rowlinson} A, et~al.
\newblock \textit{arXiv e-prints} :arXiv:2312.04237 (2023)

\bibitem{bhardwaj23}
{Bhardwaj} M, et~al.
\newblock \textit{arXiv e-prints} :arXiv:2306.00948 (2023)

\bibitem{smartt23}
{Smartt} SJ, et~al.
\newblock \textit{arXiv e-prints} :arXiv:2309.11340 (2023)

\bibitem{radice23}
{Radice} D, et~al.
\newblock \textit{arXiv e-prints} :arXiv:2309.15195 (2023)

\end{thebibliography}

\end{document}